\documentclass[aps,showkeys,nofootinbib]{revtex4-1}
\usepackage{float}
\usepackage{multirow}
\usepackage{amsmath}
\usepackage{amssymb}
\usepackage{graphicx}
\usepackage{stackrel}
\usepackage{natbib}
\usepackage{comment}
\usepackage[unicode=true]{hyperref}
\usepackage{color}
\usepackage{graphicx}

\begin{document}
	\title{Bounds on Dark Matter Annihilation Cross-sections from Inert Doublet Model in the context of 21 cm Cosmology of Dark Ages.}

	\author{Rupa Basu}
	\email{rupabasu.in@gmail.com}
	\affiliation{Department of Physics, St. Xavier's College, 
30, Mother Teresa Sarani, Kolkata-700016, India.}

\author{Madhurima Pandey}
	\email{madhurima.pandey@saha.ac.in}
	\affiliation{Astroparticle Physics and Cosmology Division, \,
Saha Institute of Nuclear Physics, HBNI\\1/AF Bidhannagar, Kolkata-700064, 
India.}
 
	\author{Debasish Majumdar}
	\email{debasish.majumdar@saha.ac.in}
	\affiliation{Astroparticle Physics and Cosmology Division, 
Saha Institute of Nuclear Physics, HBNI\\1/AF Bidhannagar, Kolkata-700064, 
India.}

	\author{Shibaji Banerjee}
	\email{shiva@sxccal.edu}
	\affiliation{Department of Physics, St. Xavier's \,College, 
30, Mother Teresa Sarani, Kolkata-700016, India.}
	
	\date{}
%
%

\begin{abstract}
We study the fluctuations in the brightness temperature of 21-cm signal $\delta T_{21}$  at the dark ages ($z\sim100$) with a dark matter candidate in Inter Doublet Model (IDM). We then explore the effects of different fractions of IDM dark matter on $\delta T_{21}$ signal. The IDM dark matter masses are chosen in few tens of GeV region as well as in the high mass region beyond 500 GeV. It has been observed that the  $\delta T_{21}$ signal is more sensitive in the dark matter mass range
of $70 - 80$ GeV. A lower bound on annihilation cross-section for this dark matter is also obtained analyzing the $\delta T_{21}$ signal.
This is found to lie within the range $6.5 \times 10^{-29} \,\, \rm{cm^3 / sec} \leq \langle\sigma v\rangle \leq 4.88\times 10^{-26}\,\, \rm{cm^ 3 / sec}$  for the IDM dark matter mass range $10 \, {\rm GeV} \leq m_{\chi}\leq 990 \,{\rm GeV} $.

\end{abstract}

\maketitle


\section{Introduction}\label{sec:1}
The cosmic dark ages that span between the era of recombination leading to thermal decoupling of baryons from photons at  $z\simeq 1100$ upto the era of re-ionization with ignition of first stars at around $z\sim 20$ could serve as a sensitive probe for any source of energy input into the cosmos. After recombination, the CMB temperature and the matter temperature remain in the same order due to Compton scattering of the residual electrons. But after $z\sim 150$ due to adiabatic expansion, the matter temperature $T_b$ goes as $T_b \propto (1+z)^2$. In case $T_b$ deviates from this dependence, it would indicate the presence of other sources of energy injection (or absorption). The measurement of brightness temperature of 21-cm hydrogen line could be such a probe \cite{Pritchard08, pritchard12, zaldarriaga04}. 

The 21-cm hydrogen line originates due to the hyperfine transition between 
triplet ($S=1$) and singlet ($S=0$) states in hydrogen atom. The triplet state manifests when the nuclear spin and the electron spin of the hydrogen atom are aligned while, the singlet state corresponds to the state when two spins are antiparallel. The strength of the transition is defined in terms of the populations of triplet and singlet states ($n_1$ and $n_0$ respectively) and the spin temperature $T_S$ of the baryon gas \cite{field58} can be described in terms of the ratio of $n_1$ and $n_0$ as
\begin{eqnarray}
\frac {n_1} {n_0} &=& 3\exp\left (- \frac {T_*} {T_S} \right )
\simeq 3 \left ( 1 - \frac {T_*} {T_S} \right )\,\, ,
\label{eq-1}
\end{eqnarray}
where $T_* = 0.068$ K represents the energy corresponding to the 21-cm transition (energy difference between the spin states $S=0$ and $S=1$ of ground state hydrogen). At $z \sim 20$ when the first star ignites, the UV radiation from this ignition initiates transitions from triplet to singlet state of hydrogen through Wouthuysen-Field effect \cite{hirata06} and the spin temperature tends to be equal to the baryon temperature. The temperature of the 21-cm line redshifted to today is defined as the brightness temperature of 21-cm line \cite{Pritchard08, pritchard12, zaldarriaga04} in the background of CMB, given by 
\begin{eqnarray}
\delta T_{21}(z) \approx  23\,\left(1 - \frac{T_\gamma}{T_{S}} \right)\,\left( \frac{\Omega_b h^2}{0.02}\right)\left(\frac{0.15}{\Omega_m h^2}\right)^{1/2} \sqrt{\frac{1+z}{10}} \, x_{\rm HI}\,\rm{mK}.
\label{eq-2}
\end{eqnarray}
where $T_\gamma$ is the CMB temperature and $x_{\rm HI}$ is the number fraction of neutral hydrogen. The spin temperature $T_S$ is smaller than the CMB temperature during dark ages. So the brightness temperature of 21-cm line is of negative value at this epoch (Eq.~(\ref{eq-2})). As an example, at $z \sim 20$ (at reionization epoch) the brightness temperature of 21-cm line can be computed as $\delta T_{21} \sim -209$ mK. The EDGES \cite{edges} experiment explored the 21-cm absorption features during the reionization era.  They reported brightness temperature of $\delta T_{21} = -500^{+200}_{-500}$ mK at 78.2 MHz frequency at the same epoch. This observed unexpected cooling may be either due to the cooler baryon temperature than expected or there are other 21-cm sources. We adopt the first possibility in this work considering the proposition that, the baryon cooling is caused by the interaction of dark matter (DM) with the baryons. In addition, DM annihilation can inject energy into the system 
\cite{furlanetto06, natarajan09} thereby influences the features of $\delta T_{21}$ brightness.

In this work, we choose the particle dark matter candidate by considering
a well established particle dark matter model namely Inert Doublet 
Model (IDM) \cite{lopez10, banerjee19, kalinowski18} in which the Standard Model (SM) of particle physics is minimally extended by adding an extra doublet scalar. With this particle candidate for dark matter we compute the heating effects due to DM-baryon interaction relevant for calculation of $\delta T_{21}$ brightness temperature by solving a set of coupled differential equations. In addition, the heat exchange due to the annihilation of these DM particles is studied in the context of the brightness temperature $\delta T_{21}$ \cite{valdes07}. In a number of literatures the viability of IDM to be a dark matter candidate have been addressed \cite{banerjee19, kalinowski18}.

In the IDM model an extra inert SU(2) doublet scalar (inert Higgs) is added with the Standard Model. 
This additional scalar SU(2) doublet does not acquire any vacuum expectation value (vev) on spontaneous symmetry breaking (SSB). A discrete Z$_2$ symmetry is imposed on the added SU(2) doublet scalar such that the added doublet is Z$_2$ even while the SM is Z$_2$ odd. Thus it can neither decay to SM fermions nor it can induce any mass to SM fermions (no vev). The stability of the added scalar is ensured by this Z$_2$ symmetry and the interaction between dark matter candidates with the SM sector is only through the Higgs portal. The lighter of the two neutral scalars of this inert doublet is then a viable candidate for dark matter. The model contains two doublet scalars. One is the usual SM Higgs doublet $H_1$ and the other is the inert doublet $H_2$. Under the imposed discrete Z$_2$ symmetry $H_2 \rightarrow H_2$. The two SU(2) scalar doublets $H_1$ and $H_2$ can be represented by 
\begin{equation}
H_1 = \left ( \begin{array}{c} \phi^+ \\
              h' + i\chi \end{array} \right )
\,\,\,\,\,
H_2 = \left ( \begin{array}{c} {H^+} \\
{(S+iA)/\sqrt{2}} \end{array} \right )\,\, .
\label{eq-higgs}
\end{equation}
In the above $\phi^+$ and $H^+$ are charged scalars, $h'$ and $\chi$ are 
real scalars, $S$ is a CP even scalar whereas $A$ is a pseudoscalar.
The interaction Lagrangian for this model can be written as 
\begin{eqnarray}
{\cal {L}} &\supset& m_{11}^2 H_1^\dagger H_1 + m_{22}^2 H_2^\dagger H_2 + 
\lambda_1 \left ( H_1^\dagger H_1 \right )^2 +
\lambda_2 \left ( H_2^\dagger H_2 \right )^2 + \nonumber \\
&& \lambda_3 \left ( H_1^\dagger H_1 \right ) 
\left ( H_2^\dagger H_2 \right ) + 
\lambda_4 \left ( H_2^\dagger H_1 \right ) \left ( H_1^\dagger H_2 \right )+ 
\nonumber \\
&& \frac {\lambda_5}{2}  
\left [ \left (H_2^\dagger H_1 \right )^2 + 
\left ( H_1^\dagger H_2 \right )^2 \right ]\,\, .
\label{eq-lag}
\end{eqnarray}
While $m_{11}$ and $m_{22}$ are the mass terms, $\lambda_i$s are various
coupling parameters of the model. 
The particle physics model for the dark matter candidate is the inert doublet 
model or IDM which is a Higgs portal model. The scattering and annihilation 
interactions of dark matter are mediated by Higgs (the dark sector is 
connected to visible sector (Standard Model) via the Higgs boson). The 
relevant coupling(s) with which dark matter couples to Higgs are then
$\lambda_3, \lambda_4, \lambda_5$, written as a single 
coupling, $\lambda_{L,L_1} = \frac {1}{2} (\lambda_3 + 
\lambda_4 \pm \lambda_5)$.

These parameters are constraint
using both the theoretical and experimental bounds. The theoretical
bounds include
the perturbativity bound, by which $|\lambda_i| < 4\pi$, vacuum stability 
condition (the interaction potential should be bounded from below) by 
which $\lambda_{1,2} > 0$, $\lambda_3 > -2\sqrt{\lambda_1\lambda_2}$,
$\lambda_{L,L_1} > -\sqrt{\lambda_1\lambda_2}$, 
$\lambda_3 + \lambda_4 - |\lambda_5| +
2 \sqrt{\lambda_1 \lambda_2} > 0$ and $\lambda_3 + 2\sqrt{\lambda_1\lambda_2} 
> 0$, the unitarity bound for which $\lambda_3 \pm \lambda_4 < 4\pi$, 
$\lambda_3 \pm \lambda_5 < 4\pi$, $\lambda_3 + 2\lambda_4 \pm 3\lambda_5 
< 4\pi$, $-\lambda_1 -\lambda_2 \pm \sqrt{(\lambda_1 - \lambda_2)^2 + 
\lambda_4^2} < 4\pi$, $-3\lambda_1 - 3\lambda_2 \pm 
\sqrt{9(\lambda_1 - \lambda_2)^2 + 2(\lambda_3 + \lambda_4)^2} < 4\pi$, 
$-\lambda_1 - \lambda_2 \pm \sqrt{(\lambda_1 - \lambda_2)^2 + \lambda_5^2}
< 4\pi$. The experimental constraints are obtained from the dark matter
relic density results given by Planck, the upper bounds of dark matter-nucleon 
scattering cross-sections for different dark matter masses obtained from
dark matter direct detection experiments as also from the collider bounds
such as LEP I measurements of $Z$ boson decay width ($M_S + M_A \geq M_Z$), 
invisible dark matter decay etc. 

After spontaneous symmetry breaking or SSB, the scalar $H_1$ develops a vev $v$
while $H_2$ does not develop any vev.  Then expanding the physical 
scalar around the minima, one obtains
\begin{equation}
H_1 = \left ( \begin{array}{c} \phi^+ \\
            \frac {h + v + i\chi}{\sqrt {2}} \end{array} \right )
\,\,\,\,\,
H_2 = \left ( \begin{array}{c} {H^+} \\
\frac {S+iA}{\sqrt{2}} \end{array} \right )\,\, .
\label{eq-higg2}
\end{equation}
In the above, $h$ is the physical Higgs boson. Now making a gauge 
transformation on the field $H_1$ to move to unitary gauge, one obtains
$H_1 = \left ( \begin{array}{c} 0 \\
            \frac {h + v}{\sqrt {2}} \end{array} \right )$ and 
the three Goldstone bosons are absorbed by $W^\pm$ and $Z$ bosons and these gauge 
bosons get their masses and longitudinal components. In Eq.~(\ref{eq-higgs}), 
both $S$ and $A$ could be candidates for dark matter, but here $S$
is considered to be the lighter of the two and is attributed as IDM 
dark matter candidate. 

After SSB, the relevant interaction vertices for the IDM dark matter $S$ and 
Higgs $h$ interaction are of the form $SSh$ and $SShh$ and the corresponding
couplings are now calculated as $g_{SSh} = \lambda_Lv$, 
$g_{SShh} = \frac {1} {2} \lambda_L$. The constraining of the coupling parameter
$\lambda_L$ by theoretical and experimental bounds have alredy been discussed.
In a recednt work \cite{stocker}, 
P. Stocker et al discusses the constraining of such coupling parameter in Higgs portal model and these are addressed
in respect of Planck observation of dark matter relic densities as well as
the dark matter direct detections while consideration has also been made
of collider results and constraints. In this work however, the same coupling parameters are contraints not only by the bounds given by these observational results
but also by imposing the theoretical bounds. Hence, while the present 
bound on the couplings satisfy the relic density bounds and other experimental bounds, they are also constraint by the 
theoretical bounds.

In this work, we consider the scalar $S$ as the dark matter candidate in IDM model to constitute the dark matter in the Universe. We then estimate the fluctuations in the brightness temperature of 21-cm signal when  the relic densities of IDM dark matter lie within $95\%$ confidence limit of the the Planck observational result \cite{plank20} for dark matter relic density. We also include the evolution of heat generated due to the DM annihilation \cite{natarajan09, furlanetto06, damico18, liu18} and DM-baryon elastic scattering \cite{munoz15, dvorkin14} in the dark ages and then its impact on the fluctuations of the brightness temperature $\delta T_{21}$ of 21-cm signal. In addition, we also explore a scenario that only a fraction of IDM dark matter takes part in the collision or annihilation process or both that affect the brightness temperature $\delta T_{21}$. To this end, we find the variation of $\delta T_{21}$ for different fractions of IDM dark matter.\\
The paper is organized as follows. In Section \ref{sec:2} we discuss the formalism for the evolution of baryon temperature $T_b$ and dark matter temperature $T_\chi$ and subsequently the evolution of $\delta T_{21}$. Section~\ref{sec:3} describes the calculations and results and finally in Section~\ref{sec:4}, we summarize the work.

\section{Formalism}\label{sec:2}
Recent EDGES experiment \cite{edges} detects the global 21-cm signal with some uncertainties. The fluctuation in 21-cm signal is quantified by its differential brightness temperature $\delta T_{21}$, which depends on the spin temperature $T_S$,  CMB temperature $T_{\gamma}$ (Eq.~(\ref{eq-2})). The spin temperature $T_{S}$ is the excitation temperature of 21-cm line which depends on the ratio of population number density of the two hyperfine splitting of hydrogen atom and that can be estimated using Eq.~(\ref{eq-1}). Further we approximate $T_S$ \cite{furlanetto07, field58} by neglecting the Wouthuysen-Field effect \cite{hirata06} as
\begin{equation}
T_S = \frac{(T_{\gamma}A_{10} + C_{10}T_{\star})\,T_b}{A_{10}T_b+C_{10}T_{\star}}
\label{eq-3}
\end{equation}
where $T_{\star} = hc/k\lambda_{\rm 21cm} = 0.068{\rm K}$, $A_{10} =2.85 \times 10^{-15} s^{-1}$ is the Einstein coefficient and $C_{10}$ is the collisional transition rate \cite{lewis07}. 

From Eqs.~(\ref{eq-2}) and (\ref{eq-3}), it can be noted that the fluctuation in the brightness temperature of 21-cm signal depends on the baryon temperature. Thermal evolution in the Universe evolves the baryon temperature $T_b$. In this analysis, we consider DM annihilation and DM-baryon elastic scattering as the additional	 contributions to the evolution of $T_b$, which is given by
\begin{equation}
	\begin{split}
		(1+z)\frac{{\rm d} T_b}{{\rm d} z} = \, & 2 T_b + \frac{\Gamma_c}{H(z)}
		(T_b - T_{\gamma})- \left( \frac{d E}{d V  d t} \right)_{\rm inj} \frac{1}{n_H} \frac{2 f_{\rm heat}(z)}{3\, H(z) (1+x_e+f_{\rm He})}-\frac{2 \dot{Q}_{b}}{ 3 H(z)}
		\end{split}
		\label{eq-4}		
	\end{equation}
where $H(a)$ is the Hubble parameter, $\Gamma_c$ is the Compton scattering rate, $x_e$ is the number of free electrons and $f_{He}$ is the relative number of abundance of the helium nuclei. The helium abundance is given by $f_{He}=n_{He}/n_{H}$. The quantity $(dE/dVdt)_{\rm{inj}}$ is the energy injection rate per unit volume and 
$dQ_b/dt$ is the heating rate of the baryons in their rest frame. The third term of Eq.~(\ref{eq-4}) represents the energy transfer rate due to the DM annihilation whereas the fourth term of Eq.~(\ref{eq-4}) accounts for the contribution due to the DM-baryon elastic scattering. 

In Eq.~(\ref{eq-4}), the term $\Gamma_c$ is related to the scattering between CMB photons and residual free electrons. The number density of CMB photons is much larger than the residual free electrons, which implies that the Compton scattering of CMB photons with the residual free electrons is efficient to keep the baryons in thermal equilibrium with CMB photons. Hence the Compton scattering rate $\Gamma_c$ depends on the number fraction of free electrons $x_e=\frac{n_e}{n_b}$ and this is given by
\begin{equation}
\Gamma_c = \left(\frac{8\,\sigma_T\,a_r\,T_\gamma^4}{3\,m_e}\right)\frac{x_e}{1+f_{He}+x_e} 
\label{eq-5}
\end{equation}
where $\sigma_T$ is the Thomson cross-section, $a_r$ is the radiation constant and $m_e$ is the mass of electron.

The annihilation of dark matter may induce additional effects in the evolution of the baryon temperature $T_{b}$. DM annihilations heat up the baryons and increase the baryon temperature $T_{b}$ and thus modify the 
 $\delta T_{21}$ spectrum. The heating of baryons by DM annihilation proceeds via two mechanisms. 
The first one is during the epoch of thermal decoupling from CMB and in this case DM annihilation enhances the fraction of free electrons ($x_e = n_e /n_b$) above the threshold value.  The evolution of $x_e$  is given as \cite{xe}.
\begin{equation}
(1+z) \dfrac{d x_e}{dz } = \dfrac{C}{H(z)} \left( n_H \mathcal A_B x_e^2 - 4 (1-x_e) \mathcal B_B e^{3E_0/(4T_\gamma)} \right),
\label{eq-6}
\end{equation} 
where $C$,\,$E_0$  are the Peebles factor \cite{peebles1968recombination} and the ground state energy of Hydrogen respectively. The effective recombination coefficient and the effective photoionization rate to the excited state and from the excited state respectively are  $\mathcal{A}_B(T_b, T_\gamma)$ and $\mathcal{B}_B(T_\gamma)$ \cite{ali2010ultrafast} \cite{ali2011hyrec}.
A large value of $x_e$ delays the CMB decoupling and as a result $T_{b}$ increases since the baryon has less time to cool adiabatically. The second mechanism involves injection of energy when DM annihilation directly heats up the baryons and increases $T_{b}$.  We have mentioned earlier that the third term of Eq.~(\ref{eq-4}) represents the DM annihilation contribution to the evolution of $T_{b}$.

The energy injection rate per unit volume $(dE/dVdt)_{\rm{inj}}$ is estimated by assuming that DM annihilates to standard model particles and injects energy into the Universe \cite{damico18}. This process further drives additional ionization, excitation and heating of the gas. For a given velocity averaged annihilation cross section $\langle\sigma v\rangle$, $dE/(dVdt)_{\rm{inj}}$ is given by
\begin{equation}
\left(\frac{dE}{dVdt} \right)_{\rm{inj}}= \rho^{2}_{\chi}\,B(z)\,f^{2}_{\chi}\frac{\langle\sigma v\rangle}{M_{\chi}}
\label{eq-7}
\end{equation}
where $f_{\chi}$ is the fraction of the dark matter that annihilates to standard model particles and $\rho_{\chi}$ is the dark matter density. In the above, we consider two different expressions for $B(z)$, the boost factor. These are $B(z) = 1 + 1.6 \times 10^5 a^{1.54}\,{\rm Erfc}(\frac{1+z}{20.5})$ and $B(z) = 1 + 2.3 \times 10^6 a^{1.48}\,{\rm Erfc}(\frac{1+z}{19.6})$. The boost factor is a parameter which is related to the structure formation. The DM annihilations effectively occur in many small overdensities. Since the spatial average of the annihilation rate depends on the average of the square of dark matter densities, the boost factor enhances this dependence \cite{poulin2015dark}.

 The injected energy is deposited in the baryons mainly in three different ways, namely ionization, excitation and heating \cite{damico18}. The dimensionless quantity $f_{heat}$ in Eq.~(\ref{eq-4}) represents the efficiency of energy deposition in the baryon by heating. The quantity $f_{heat}$ depends on the DM mass and this includes the time delay between the energy deposition and injection of energy. In our work,
the instantaneous deposition of energy refers to the event that the fraction of the energy produced by DM annihilation at a certain redshift is immediately transferred to the gas/background. For delayed deposition however, the energy deposited to the gas/background from the DM annihilation at a certain redshift includes the delayed transfer functions given in Ref. \cite{slatyer16a}. In this analysis we estimate $f_{heat}$ following the reference \cite{liu18,slatyer16a,slatyer16b} and the analysis is performed with the so called SSCK approximation \cite{damico18}. The quantity $f_{heat}$,  is given by 
\begin{equation}
f_{heat}=f_{\rm eff}\left(\frac{1+2\,x_e}{3}\right)
\label{eq-8}
\end{equation}
where $f_{\rm eff}$ is the fraction of the energy produced by DM annihilation immediately transferred to the plasma. Following the Refs. \cite{slatyer16a, slatyer16b}, the values of $f_{\rm eff}$ are adopted\footnote{https://faun.rc.fas.harvard.edu/epsilon/} for the photons and the $e^+ e^-$ pairs injected at keV-TeV energies. Although it has been discussed in Ref. \cite{slatyer16a} that the efficiency factor $f(z)$ at a redshift $z$ (ratio of deposited and injected power) can be approximated by the value of $f_{\rm eff}$ for WIMP dark matter, but in our calculation we consider  $f_{\rm eff}$ to depend on $z$\cite{liu18,slatyer16a,slatyer16b,slatyer2009cmb}.

 The elastic scattering between DM and baryons affects the baryon temperature $T_{b}$ and its evolution. This interaction process 
may cool or heat the baryons. In general during the interaction between two fluids (DM and baryons, say), the hot fluid losees its energy to the colder one and there will be no transfer of energy if both  the fluids are at same temperature. According to  Ref. \cite{munoz15}, if there exists a relative velocity between two fluids then there should be an additional term of friction which will tend to damp the motion and the consequent kinetic energy loss will increase the temperature of both the fluids. In Ref. \cite{munoz15}, it has been shown that the magnitude of this interaction effect depends on the initial relative velocity which is a Gaussian variable with variance of $\sim 29\,{\rm km\,s}^{-1}$ at $z=1010$.

We follow the methodology prescribed in Ref.~\cite{munoz15} to estimate the quantity $dQ_b/dt$ (Eq.~(\ref{eq-4})), the heating rate of the baryons in their rest frame.  The interaction between DM and baryons at different temperature heats up the cold dark matter, where the heating rate is proportional to the temperature difference. However if there is a relative velocity between DM and baryons then the friction term heats up both DM and baryons irrespective of their temperature difference. Using Eq.~(16) of Ref.~\cite{munoz15} the heating rate of baryon is given by

\begin{equation}
\frac{dQ_{b}}{dt}=\dfrac{2\, m_b\, \rho_{\chi}\, \sigma_0\, e^{-\frac{r^2}{2}}\left(T_{\chi}-T_b\right)}{(m_{\chi}+m_b)^2 \sqrt{2\pi}\,  u_{th}^3} +\frac{\rho_{\chi}}{\rho_m} \frac{m_{\chi}m_b}{m_{\chi}+m_b}\,V_{\chi b} \left(\frac{dV_{\chi b}}{dt}\right)
\label{eq-9}
\end{equation}
where $m_{\chi}$ and $m_b$ are the mass of DM and baryon respectively. $\rho_{\chi}$ is the energy density of DM, $\rho_{b}$ is the energy density of baryon and the total matter density $\rho_{m} = \rho_{b}+\rho_{\chi}$. The term $u_{th}$ is the variance of the relative velocity between DM and baryon and this is estimated as $u_{th}\equiv \sqrt{T_b/m_b+T_{\chi}/m_{\chi}}$.
Following Ref.~\cite{munoz15} we consider that the interaction cross section 
$\sigma$ is parameterized as $\sigma=\sigma_0 v^{-4}$. We note that $\sigma_0$ depends on the DM mass $m_{\chi}$ and this is scaled as $\sigma_0 = (m_{\chi}/{\rm GeV})\times10^{-42}\,{\rm cm^{-2}}$. The quantity $V_{\chi b}$ in Eq.~(\ref{eq-9}) is the drag term of the relative velocity between DM and baryon. 

\section{Calculations and Results}\label{sec:3}
As discussed earlier, we have adopted a particle dark matter
candidate in the framework of inert doublet model or IDM and explore it's
contribution to the evolution of 21cm absorption temperature during 
dark ages and during the ionization epoch. The dark matter-baryon 
scattering and dark matter annihilation induce heat exchange of 
the baryons with the background and thus influence the evolution 
of the temperature of 21cm absorption line. 

The DM annihilation produces heating effects on baryons and directly
injects energy to the baryons and as a result, baryon temperature 
$T_b$ increases. This affects the spin temperature $T_S$ and consequently
$\delta T_{21}$. DM-baryon elastic scattering heats the baryon due to the relative velocity between them and the baryon temperature $T_b$ increases, which further 
modifies $\delta T_{21}$. We also  have estimated the effects on $\delta T_{21}$ in case a fraction of the present IDM dark matter takes part in the above processes. 

\begin{table}[h]
\caption{The relic density $(\Omega_c)$ and the corresponding annihilation cross section $(\langle\sigma v\rangle)$ for different IDM dark matter masses $(m_{\chi})$ for which the values of $\Omega_c$ lie within $95\%$ confidence limit of $\Omega_{c,0}$ obtained from the Planck experiment.}
{\begin{tabular}{@{}ccc@{}} \toprule
$m_\chi$ & $\Omega_c$ & $\langle\sigma v\rangle$ \\
(GeV)&  & ${\rm cm^{3}\,s^{-1}}$ \\ \colrule
$10$\hphantom{00} & \hphantom{00}$0.113$ & \hphantom{00}$6.50\times 10^{-29}$  \\ 
 \hline
 $20$\hphantom{00} & \hphantom{00}$0.116$ & \hphantom{00}$7.72\times 10^{-29}$  \\
 \hline
 $30$\hphantom{00} & \hphantom{00}$0.114$ & \hphantom{00}$9.98\times 10^{-29}$   \\
 \hline
 $40$\hphantom{00} & \hphantom{00}$0.116$ & \hphantom{00}$1.65\times 10^{-28}$   \\
 \hline
 $50$\hphantom{00} & \hphantom{00}$0.115$ & \hphantom{00}$4.37\times 10^{-28}$   \\
 \hline
  $60$\hphantom{00} & \hphantom{00}$0.116$ & \hphantom{00}$1.40\times 10^{-27}$   \\ 
 \hline
  $70$\hphantom{00} & \hphantom{00}$0.119$ & \hphantom{00}$1.72\times 10^{-26}$  \\ 
 \hline
  $80$\hphantom{00} & \hphantom{00}$0.113$ & \hphantom{00}$2.59\times 10^{-26}$  \\ 
 \hline
  $550$\hphantom{00} & \hphantom{00}$0.115$ & \hphantom{00}$6.68\times 10^{-26}$   \\ 
 \hline
  $990$\hphantom{00} & \hphantom{00}$0.113$ & \hphantom{00}$4.88\times 10^{-26}$ \\ \botrule
\end{tabular} }
\label{tab:1} 
\end{table}

\begin{figure}[h]
\centering
\centerline{\includegraphics[scale=0.6]{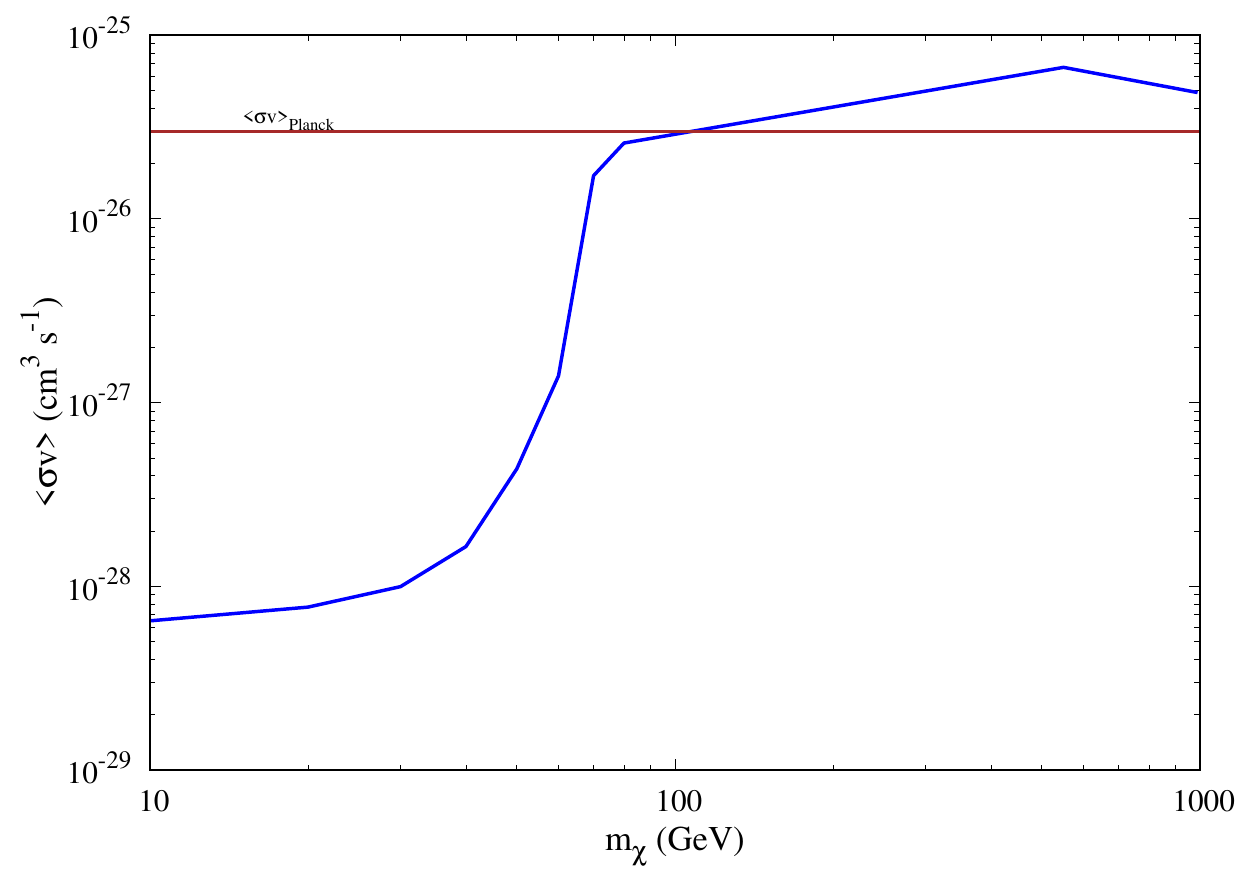}}
\caption{Annihilation cross section $\langle\sigma v\rangle$ for the different dark matter mass $m_{\chi}$. The brown curve denotes the range of $\langle\sigma v\rangle$ estimated from Planck Experiments \cite{lopez2013constraints,kawasaki2016cmb}.}
\label{fig:cross}
\end{figure}

The mass of the Higgs portal IDM dark matter is chosen to be few tens of GeV
for the calculations. But following Ref. \cite{kalinowski18,banerjee19}, where IDM dark matter of 
masses with hundreds of GeV are also adopted, we have added in our analysis 
IDM candidates of higher masses ($\gtrsim 500$ GeV) \cite{banerjee19} also. The relic densities
of the IDM dark matter candidates with chosen masses are then calculated
using the available code microOMEGAS \cite{micromega} and compared with dark matter 
relic densities given by the analysis of Planck observational results \cite{plank20}.
The relic densities and corresponding dark matter annihilation cross-sections
for each of the dark matter masses chosen for the calculations are then 
adopted for which the calculated relic densities are within the 95\% confidence limits (C.L.)
obtained from the Planck relic density results for dark matter. These 
are tabulated in Table \ref{tab:1} for the chosen dark matter masses $m_\chi$. The 
results are also plotted in Fig.~\ref{fig:cross}. In Fig.~\ref{fig:cross} the IDM dark matter 
annihilation cross-sections are shown for various IDM masses that satisfy 
the Planck relic density results within it's 95\% C.L.

\begin{figure}[h]
\centering
\centerline{\includegraphics[scale=0.6]{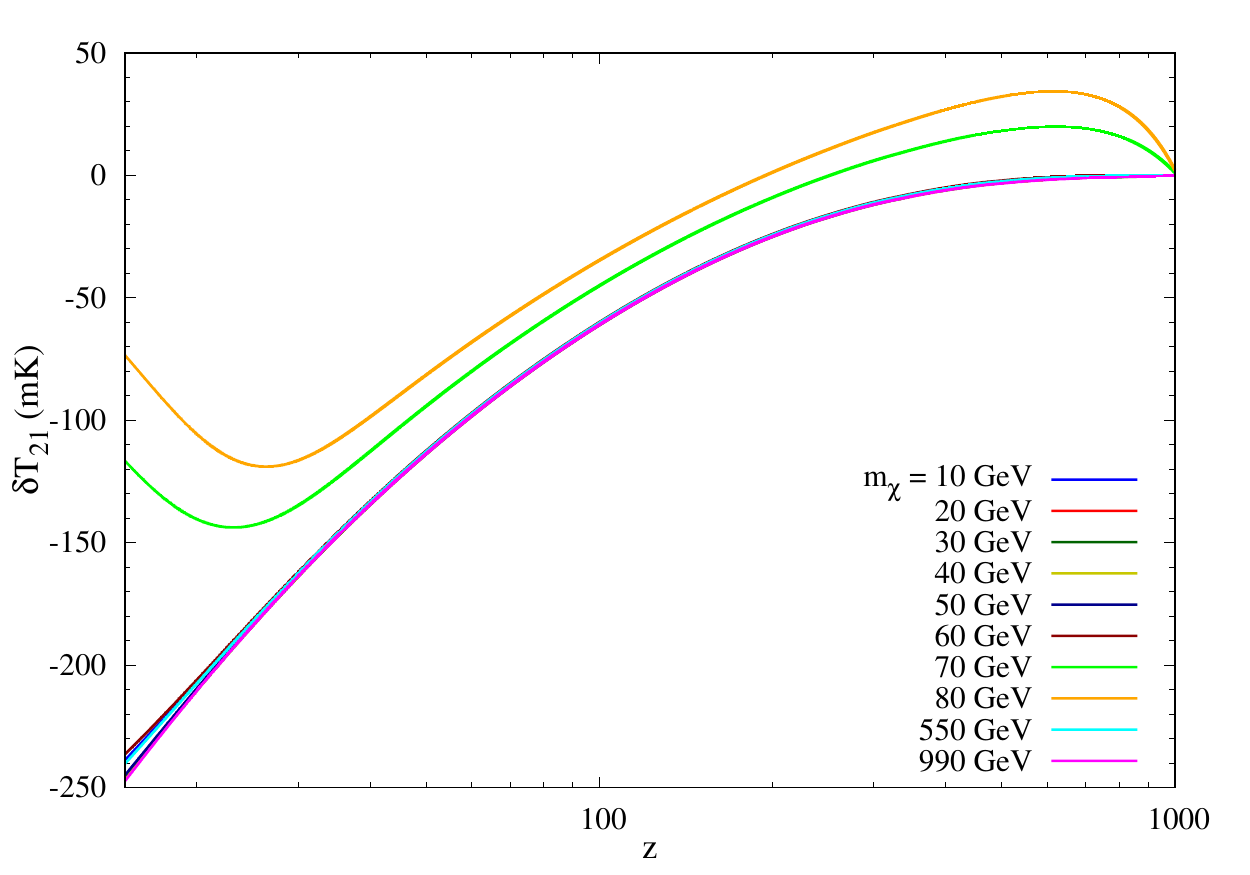}
\includegraphics[scale=0.6]{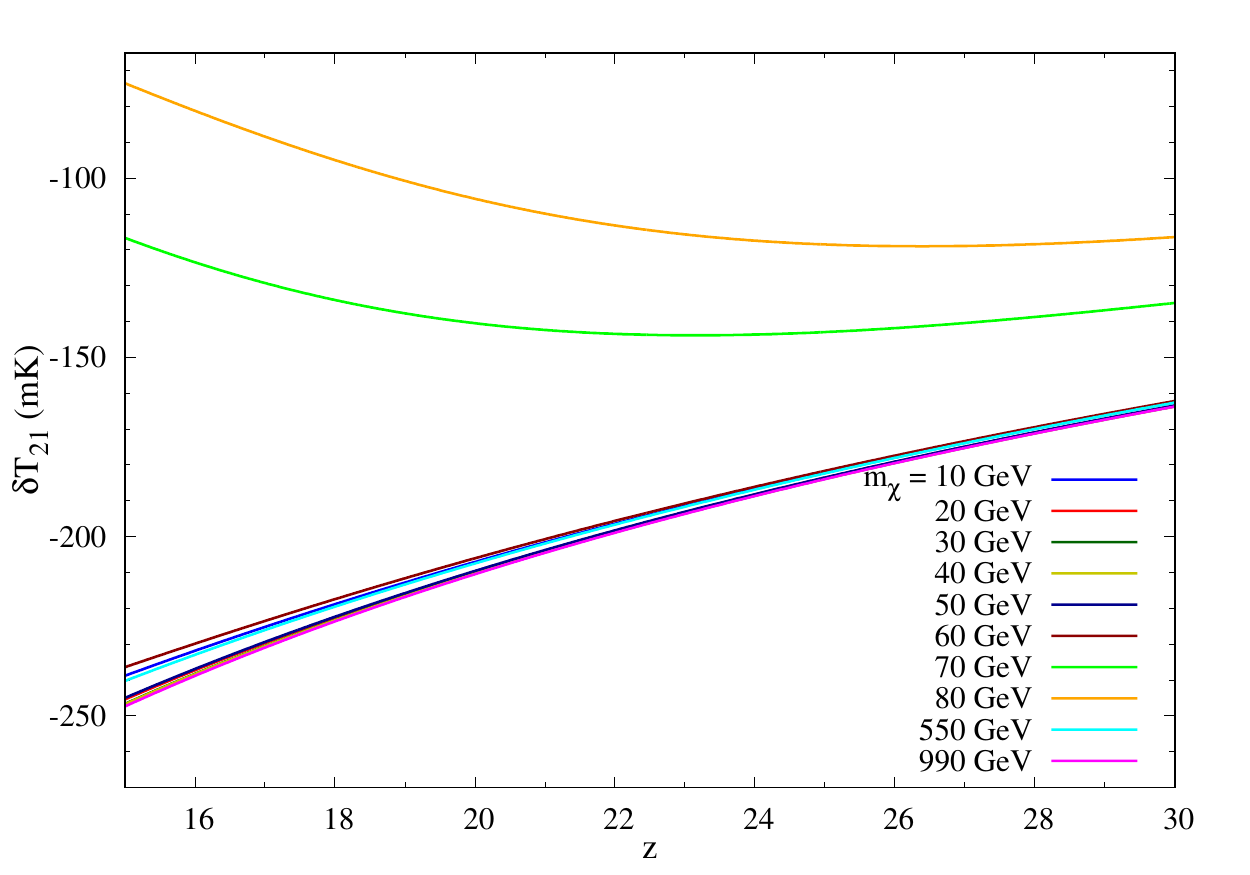}}
\caption{The fluctuation in the 21-cm brightness temperature $\delta T_{21}$ at different redshift by considering the DM annihilation as an additional effect to $\delta T_{21}$ apart from the thermal evolution. Different color lines corresponds to different IDM mass mentioned in the figure.}
\label{fig:ani}
\end{figure}

We first investigate the fluctuations in the 21cm brightness temperature
$\delta T_{21}$ by including only the dark matter annihilation contribution 
to the
usual temperature evolution of baryon temperature ($T_b$). 
For this purpose,
we consider Eq.~(\ref{eq-4}) without the last term on right hand side (RHS) of Eq.~(\ref{eq-4})
but keeping the term involving $\left (\frac {dE} {dV dt} \right )_{\rm inj}$ as the 
latter is related to the heat injection due to dark matter annihilation 
(Eq.~(\ref{eq-7})) while  the former term with ${\dot {Q}_b}$ involves the heating rate due to dark matter-baryon collision (Eq.~(\ref{eq-9})). Eqs.~(\ref{eq-4}-\ref{eq-8}) are then solved simultaneously and $T_b$ with $T_\gamma(z) = T_\gamma ^0 \,(1+z)$ (where $T_\gamma (z)$ is the background CMB temperature at a redshift $z$ and $T_\gamma^0$ is the same at the present rpoch), the spin temperature $T_S$ and hence $\delta T_{21}$ are then computed using Eq.~(\ref{eq-3}) and Eq.~(\ref{eq-2}) respectively. The values of baryon density parameter $\Omega_b (= \frac{\rho_b}{\rho_c},\, \rho_c$ being critical density) and matter density parameter $\Omega_m (= \frac{\rho_m}{\rho_c})$ are taken to be $\Omega_b = 0.04\, (1+z)^3$ and $\Omega_m = \Omega_\chi + \Omega_b = 0.30 \,(1+z)^3 $
 at different redshifts for each of the chosen dark 
masses. The results are plotted 
in the left panel of Fig.~\ref{fig:ani}. The right panel of Fig.~\ref{fig:ani} is the magnified 
representation of the left panel within the truncated redshift range 
of $z \sim 15-30$, which broadly represents the reionization epoch.  

\begin{figure}[h]
\centerline{\includegraphics[scale=0.6]{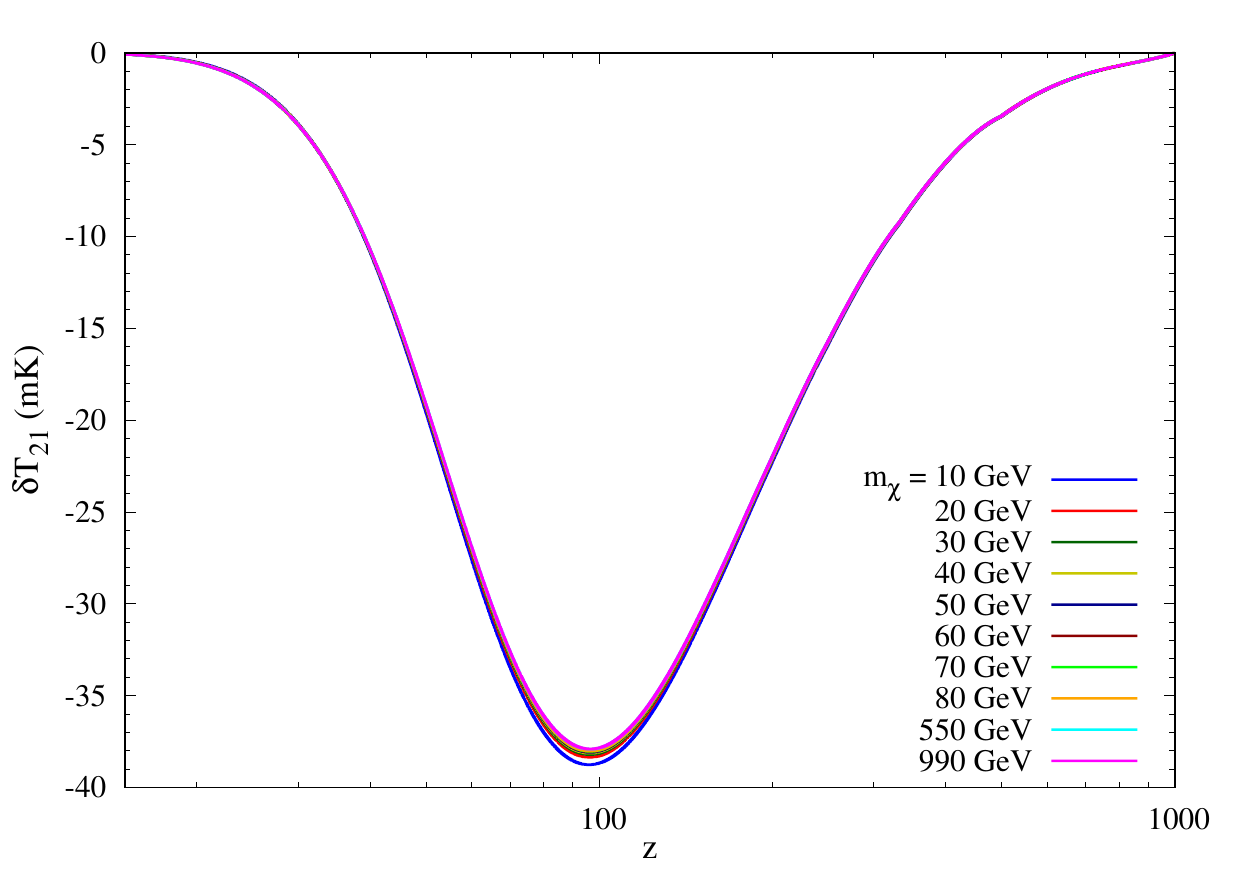}
\includegraphics[scale=0.6]{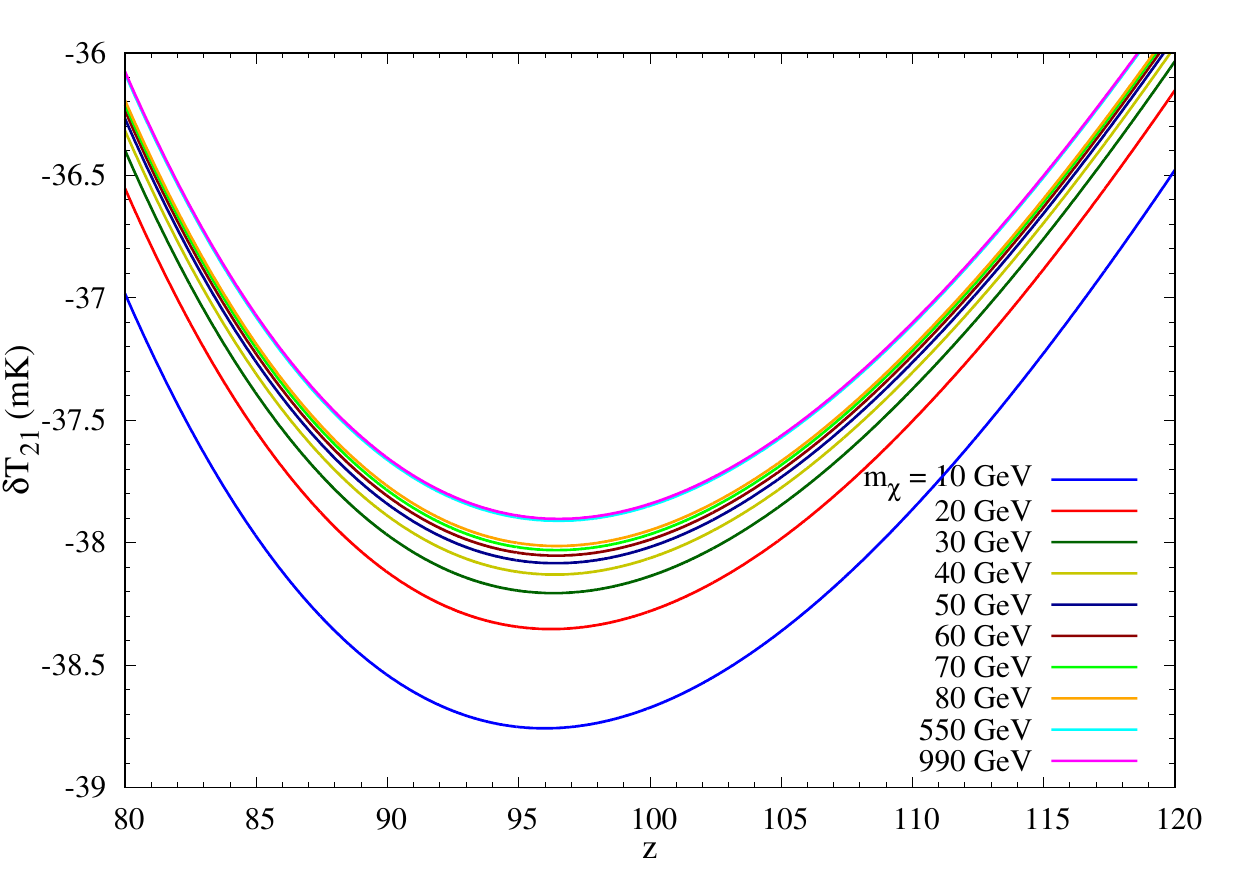}}

\caption{The fluctuation in the 21-cm brightness temperature $\delta T_{21}$ at different redshift by considering the DM-baryon elastic scattering as an additional effect to $\delta T_{21}$ apart from the thermal evolution. Different color lines corresponds to different IDM mass mentioned in the figure.}
\label{fig:col}
\end{figure}

From Fig.~\ref{fig:ani} it is observed that when the IDM dark matter mass
is in the range $\sim 70$ GeV to $\sim 80$ GeV, a dip or minima 
occurs for $\delta T_{21}$
temperature around $z \sim 21$ (for every chosen mass) which is in the region of reionization 
epoch. But for other masses greater or less than $\sim 60$ GeV, $\delta T_{21}$
is as low as $\sim -250$ mK at redshift $z \sim 10$ GeV. At redshift
$z \sim 1000$, the $\delta T_{21}$ temperature is $\sim 0$ mK for all the 
IDM dark matter masses ranging from 10 GeV to 990 GeV. The redshift $z \sim 1000$
is however the epoch when the photons become free (CMB) and is the onset
of dark ages. The difference in the nature of variations of $\delta T_{21}$  
in the mass range of  $\sim 70$ GeV to $\sim 80$ GeV throughout the dark ages
leading to the reionization epoch and when the dark matter masses
$< 70$ GeV and $> 80$ GeV  may be attributed to the fact that 
IDM dark matter is a Higgs portal dark matter and the resonance in 
their interaction cross-sections occur in the ballpark of $\sim$ 70 GeV. \\

\begin{figure}[h]
\centerline{
\begin{tabular}{@{}c c@{}}
\includegraphics[width=.48\textwidth]{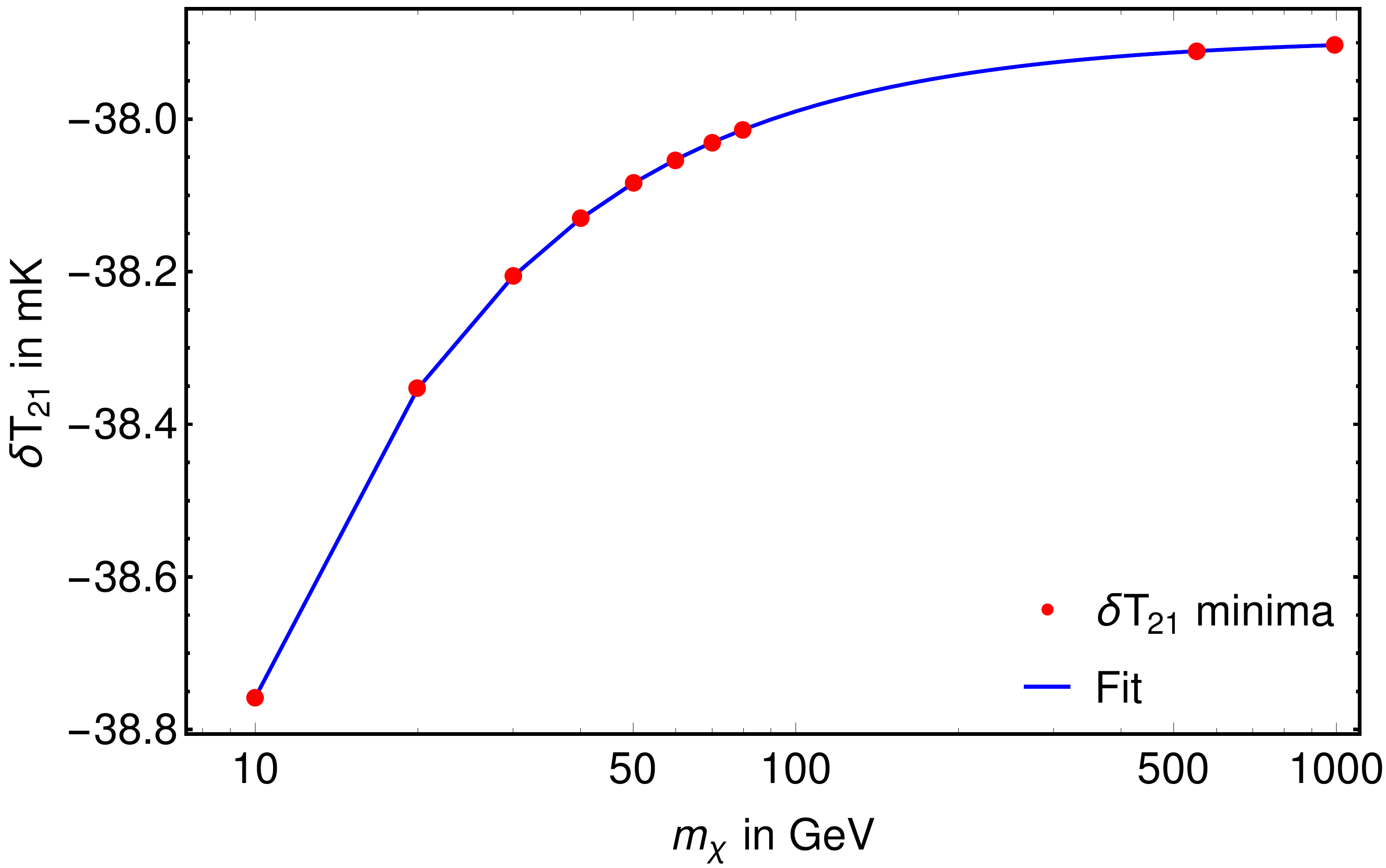}	&
\includegraphics[width=.47\textwidth]{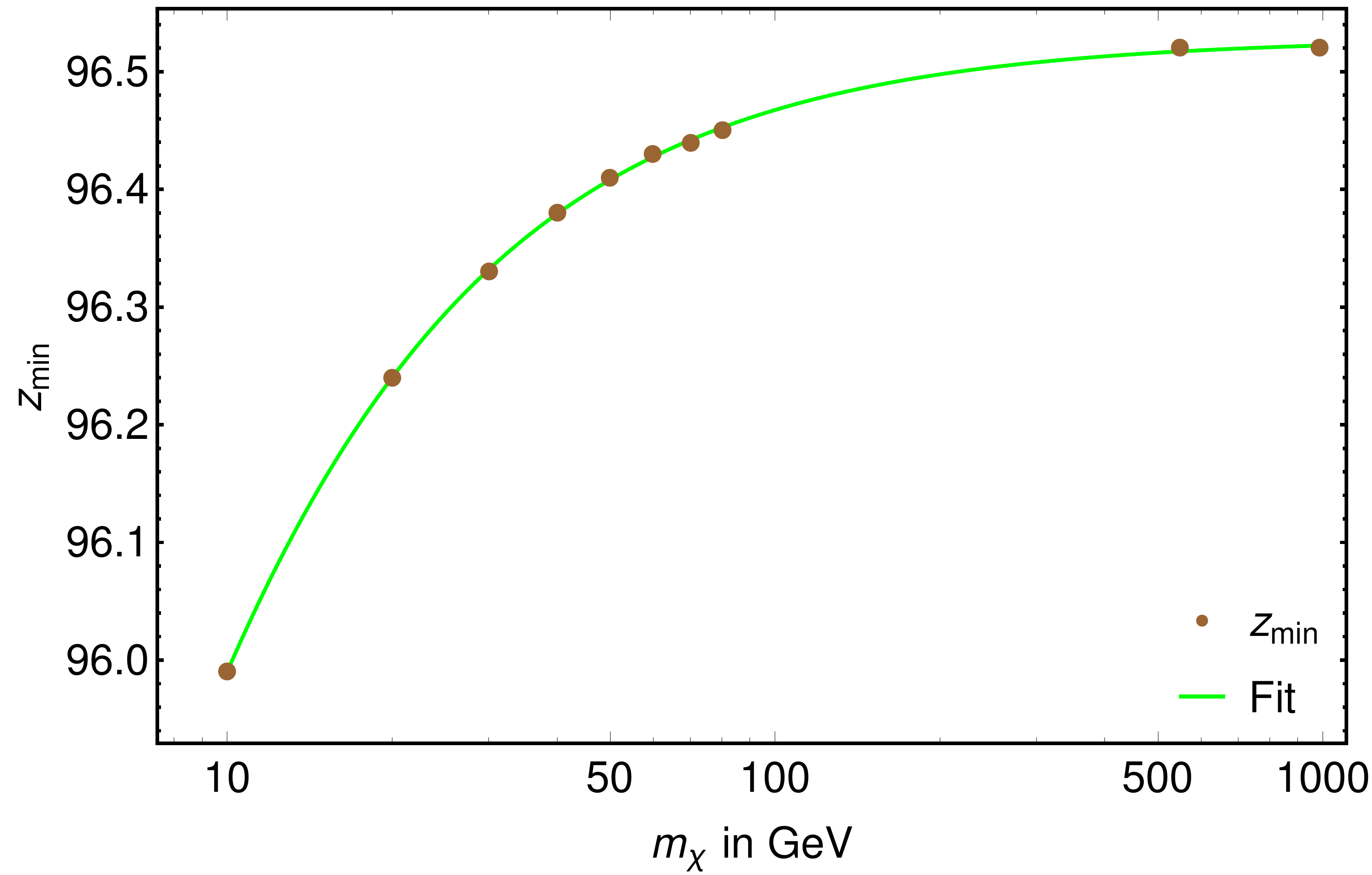} \\
(a) & (b) \\
\end{tabular}}
\caption{(a) The fluctuation in the 21-cm brightness temperature $\delta T_{21}$ at different redshift by considering the DM-baryon elastic scattering as an additional effect to $\delta T_{21}$ apart from the thermal evolution (red points) along with the fitted curve (blue line). (b) same as (a) but for the variation of 
the redshift values ($z_{\rm min}$), corresponding to each different 
$m_\chi$, at which $\delta T_{21}$ attains minimum  for that $m_\chi$. The solid green line indicates the trends of the variation of $z_{\rm min}$.}
\label{fig:colfit}
\end{figure}

Now we consider the effects of baryon-dark matter (IDM) collision to the
evolution of baryon temperature and subsequently the evolutions of 21cm 
temperature $\delta T_{21}$ throughout the dark ages. 
In doing this we omit the 
third term on the RHS of Eq.~(\ref{eq-4}) (the term associated with the dark 
matter annihilation) but keep the last term involving ${\dot Q}_b$ 
that relates to the rate of baryon heating due to dark matter-baryon 
collision. As in the earlier case we now solve computationally the 
coupled Eqs.~(\ref{eq-4}-\ref{eq-8}) along with Eqs.~(\ref{eq-2},\ref{eq-3}) for the same set of dark matter
masses as in the previous case and obtain the variation of $\delta T_{21}$
through the dark ages for each of the chosen IDM dark matter masses. 
The results are plotted in Fig.~\ref{fig:col}. In the left panel of Fig.~\ref{fig:col} we plot
the variations of $\delta T_{21}$ with $z$ for each of the chosen dark matter 
masses in the range 10 GeV to 80 GeV as also for masses 550 GeV and 990
GeV. From Fig.~\ref{fig:col} it is observed that for each of the chosen masses there 
is a dip in $\delta T_{21}$ at redshift $z$ around 95. It also appears from the left 
panel of Fig.~\ref{fig:col} that the variations of $\delta T_{21}$ through the dark ages 
are almost degenerate for the range of IDM dark matter masses chosen 
except near the minima of the $\delta T_{21}$ at $z \sim 95$ in these variations.
For the sake of more clarity and for better understanding of the redshift
region in and around of $\delta T_{21}$ minima, the redshift region 
$80 \leq z \leq 120$ is magnified in the right panel of Fig.~\ref{fig:col}.
From the right panel of Fig.~\ref{fig:col} it can be seen that the minimum value
of $\delta T_{21}$ increases as the chosen dark matter masses increases 
from 10 GeV. In fact this shift of the minimum $\delta T_{21}$ with the increase 
of chosen dark matter masses $m_\chi$ (shown in Fig.~\ref{fig:col}) appears to fit  
well with the analytical relation $\delta T_{21}\Big|_{\rm min} = -41.76 + 3.87
\exp(-\frac {2.53} {m_\chi})$. This is shown in Fig.~\ref{fig:colfit}(a). The values of redshift $z$ ($z_{\rm min}$) at which the minimum for 
$\delta T_{21}$ is obtained for each of the chosen dark matter masses ($m_\chi$) 
are shown in Fig 4(b). It appears that  $z_{\rm min}$ also follows a similar
trend as Fig.~\ref{fig:colfit}(a). 

\begin{figure}[h]
\centerline{\includegraphics[scale=0.6]{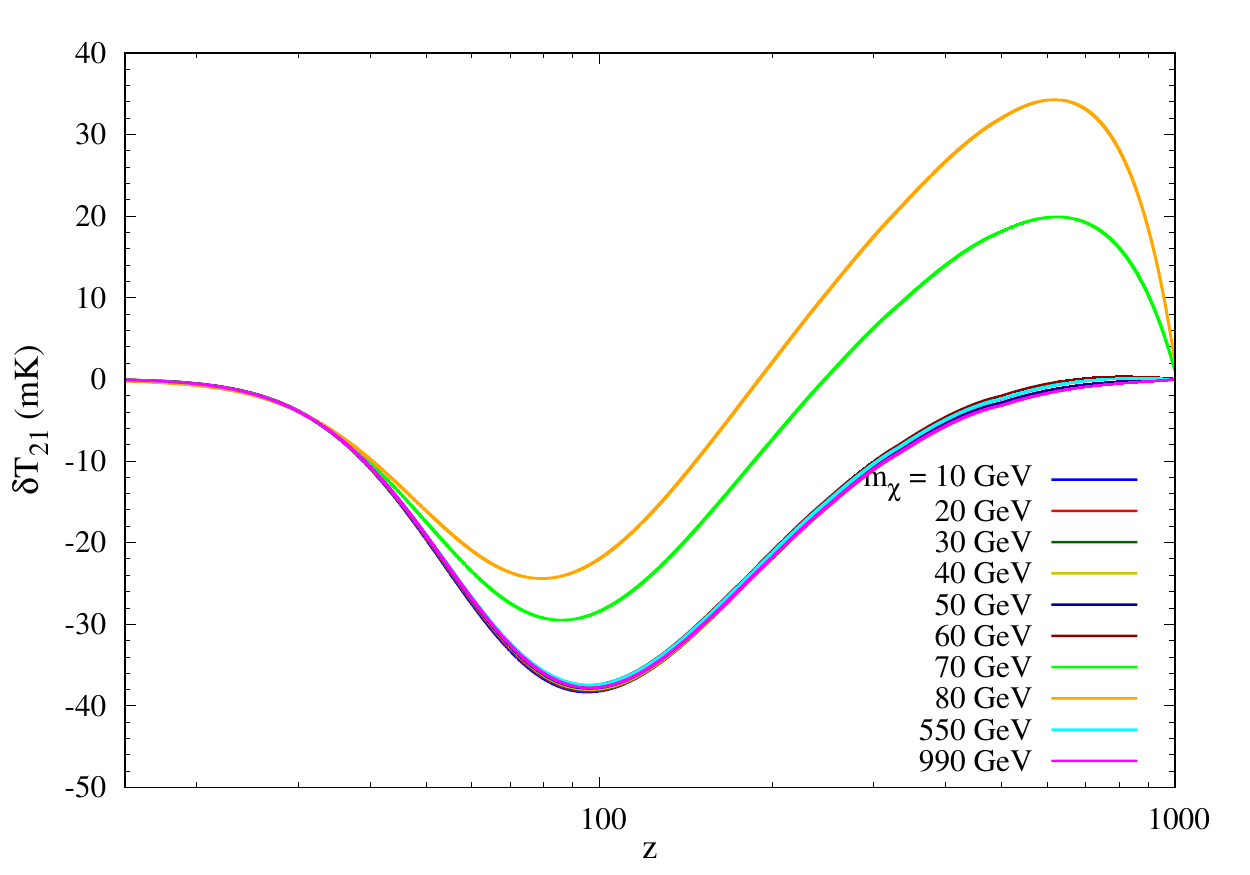}
\includegraphics[scale=0.6]{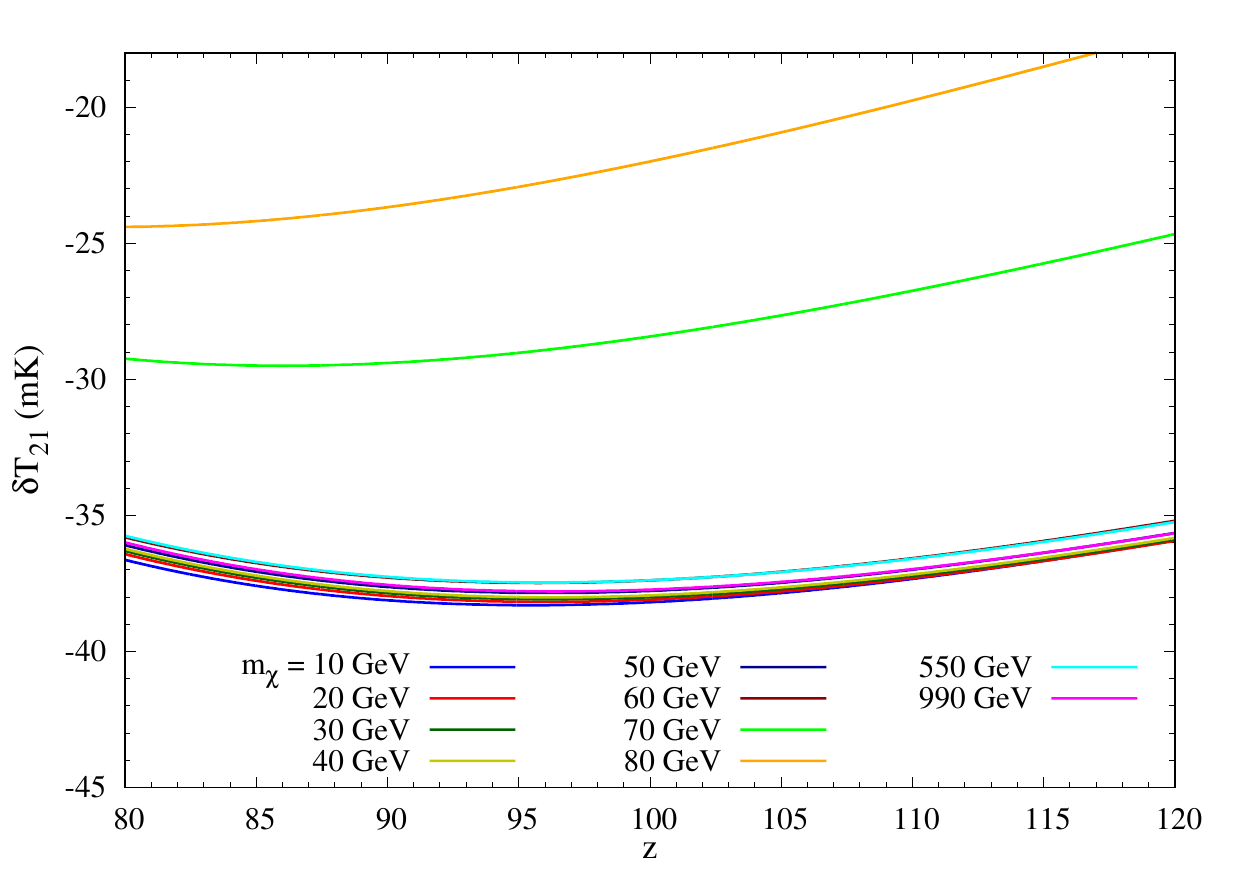}}
\caption{The fluctuation in the 21-cm absorption lines $\delta T_{21}$ (left panel) by considering both DM annihilation and DM-baryon elastic scattering as additional effects on $\delta T_{21}$ along with the thermal evolution. Different color lines correspond to different DM mass mentioned in the figure. The right panel shows the zoomed version of the figure at left panel for $80\leq z\leq 120$.}
\label{fig:total}
\end{figure}

We explore both the dark matter annihilation effects and 
dark matter-baryon collision effects together and solve numerically 
Eqs.~(\ref{eq-4}-\ref{eq-8}) keeping contributions of both these phenomena to the 
evolution equation of baryon temperature (Eq.~(\ref{eq-4})). The temperature 
$\delta T_{21}$ as a function of $z$ is then obtained using Eqs.~(\ref{eq-2}) and (\ref{eq-3}). 
The results obtained are shown in Fig.~\ref{fig:total}. In what follows, these results 
are referred to as ``combined" results. Similar to the presentations of Figs.~\ref{fig:ani} and \ref{fig:col}, the 
right panel of Fig.~\ref{fig:total} is the magnified depiction of the left panel 
of Fig.~\ref{fig:total} within a limited redshift region ($80 \leq z \leq 120$). 
From the left panel of Fig.~\ref{fig:total}, it can be seen that the low $z$ feature of $\delta T_{21}$ as seen in Fig.~\ref{fig:ani} (with only the annihilation effects) is absent. This feature is overwhelmed by the dark matter-baryon collision effect
(left panel of Fig.~\ref{fig:col}) and for the ``combined" case also the minima of $\delta T_{21}$ temperature for the chosen IDM dark matter masses hover around
$z \sim 95$. But the effects due to the annihilation term in Eq.~(\ref{eq-4}) are 
evident in the nature of variations of the minimum $\delta T_{21}$ values 
($\delta T_{21}\Big|_{\rm min}$) with different chosen masses of IDM dark matter.
While in Fig.~\ref{fig:col}, the $\delta T_{21}\Big|_{\rm min}$ gradually increases with $m_\chi$ 
following roughly an empirical relation mentioned above, in Fig.~\ref{fig:total} 
(the ``combined" result) it can be seen that the features of $\delta T_{21}$ 
for the Higgs portal IDM dark matter mass $m_\chi = 70$, 80 GeV are 
visibly different from 
the other chosen values of $m_\chi$. In fact for other masses, 
lighter or heavier than $m_\chi = 70, 80$ GeV mass range, 
$\delta T_{21}$ values are less than those for $m_\chi = 70$, 80 GeV and 
they are almost
degenerate in the redshift range $\sim 50 \leq z \leq 1000$. 
For $\sim 50 \leq z \leq 200$, baryons cool adiabatically
in the absence of any DM interaction but in the presence of DM 
annihilation and DM-baryon scattering, the
baryon temperature $T_b$ is modified accordingly affecting 
$T_S - T_\gamma$. In addition, delayed energy deposition becomes operative
at $z \sim 100$. However, beyond $z \sim 40$ although the dark matter-baryon
collision effects dominate but the quantitative nature of Fig.~\ref{fig:ani} is boardly evidence for higher $z$ values. 

\begin{figure}[h]
\centerline{
\begin{tabular}{@{}c c@{}}
\includegraphics[width=.48\textwidth]{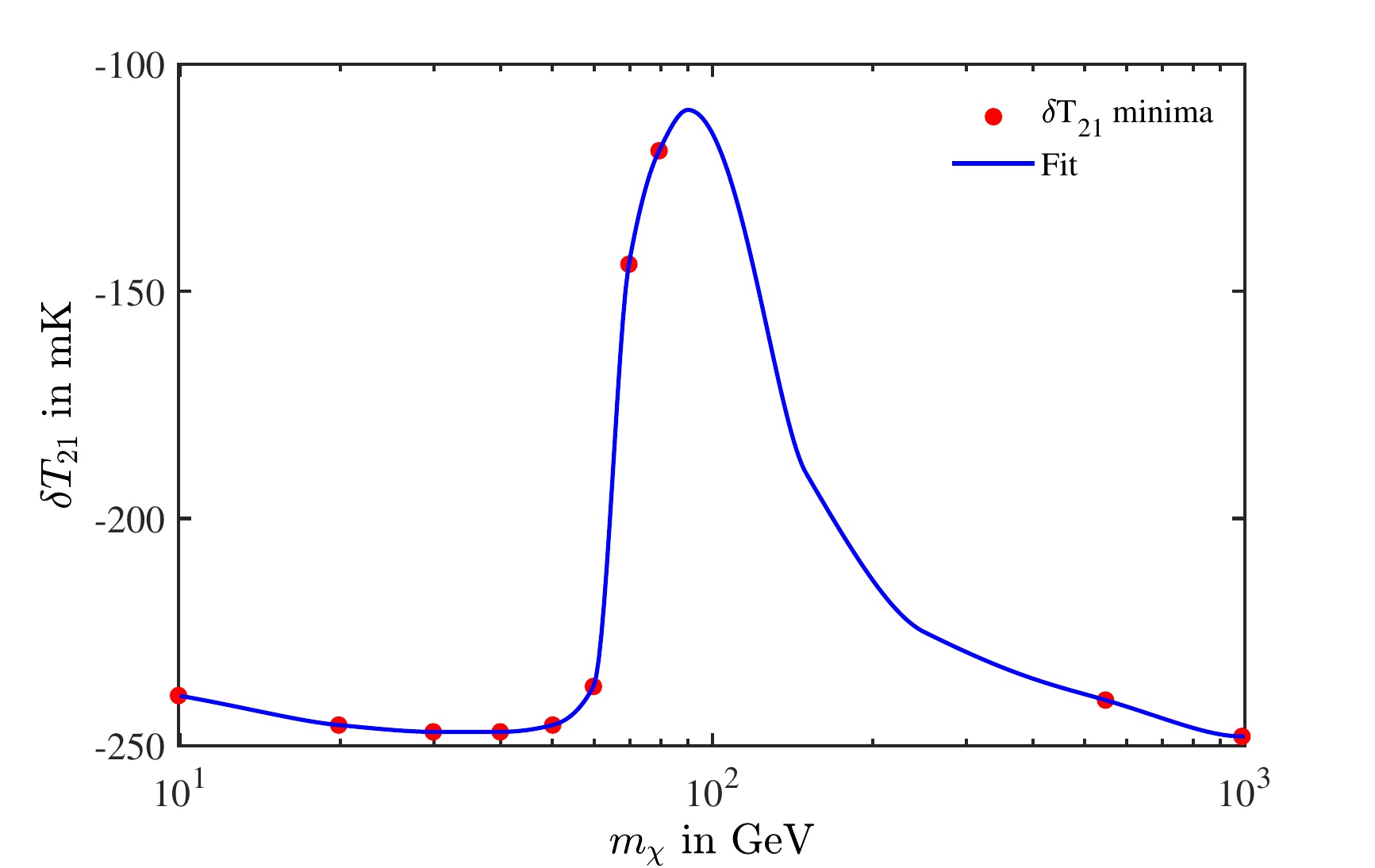}	&
\includegraphics[width=.47\textwidth]{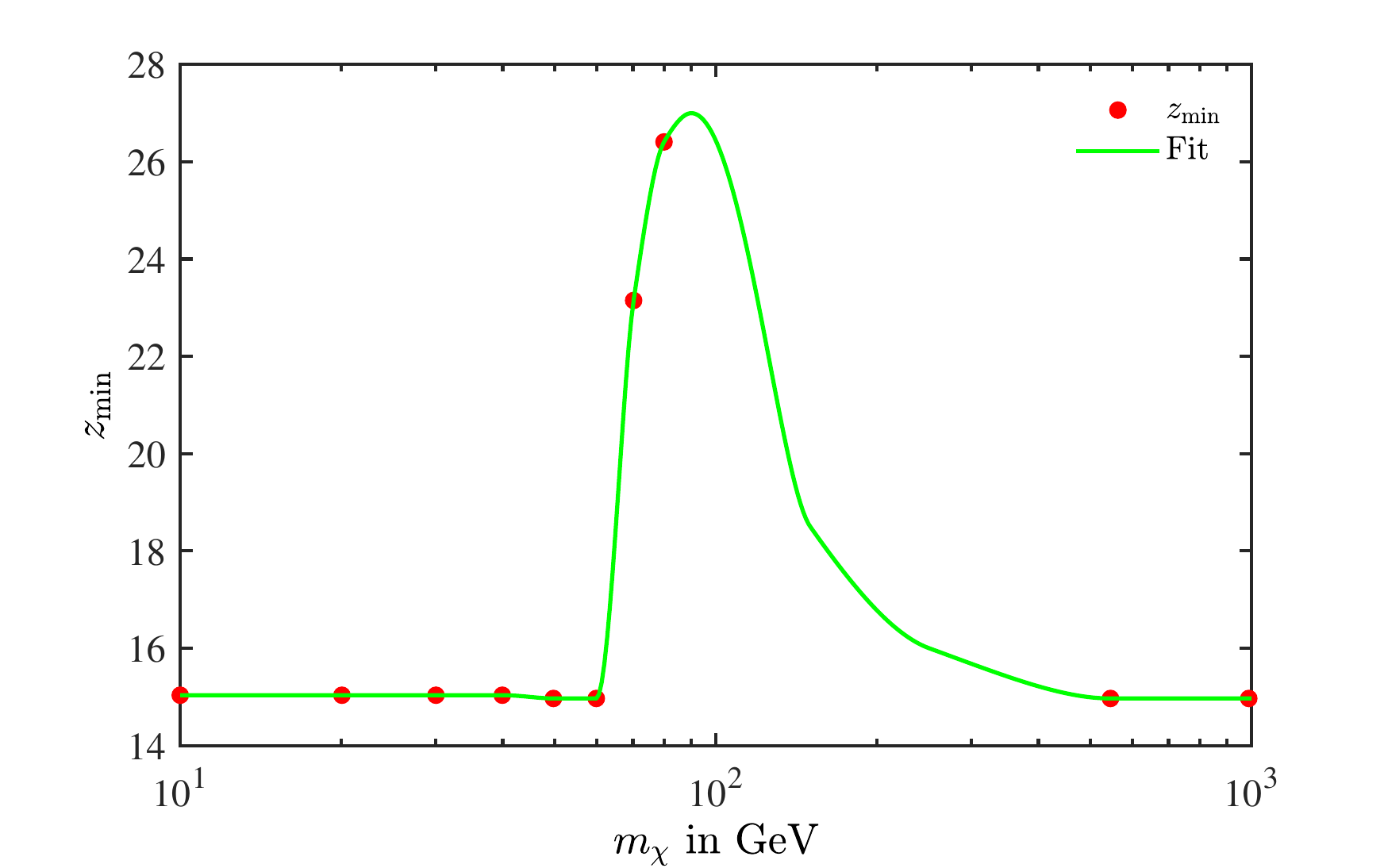} \\
(a) & (b) \\
\end{tabular}}
\caption{(a) The fluctuation in the 21-cm brightness temperature $\delta T_{21}$ at different redshift when only dark matter annihilation is the additional
effect to $\delta T_{21}$ apart from the thermal evolution (red points) along with the fitted curve (blue line). (b) same as (a) but for the variation of 
the redshift values ($z_{\rm min}$), corresponding to each different 
$m_\chi$, at which $\delta T_{21}$ attains minimum  for that $m_\chi$. The solid green line indicates the trends of the variation of $z_{\rm min}$.}
\label{fig:anifit}
\end{figure}

\begin{figure}[h]
\centerline{
\begin{tabular}{@{}c c@{}}
\includegraphics[width=.48\textwidth]{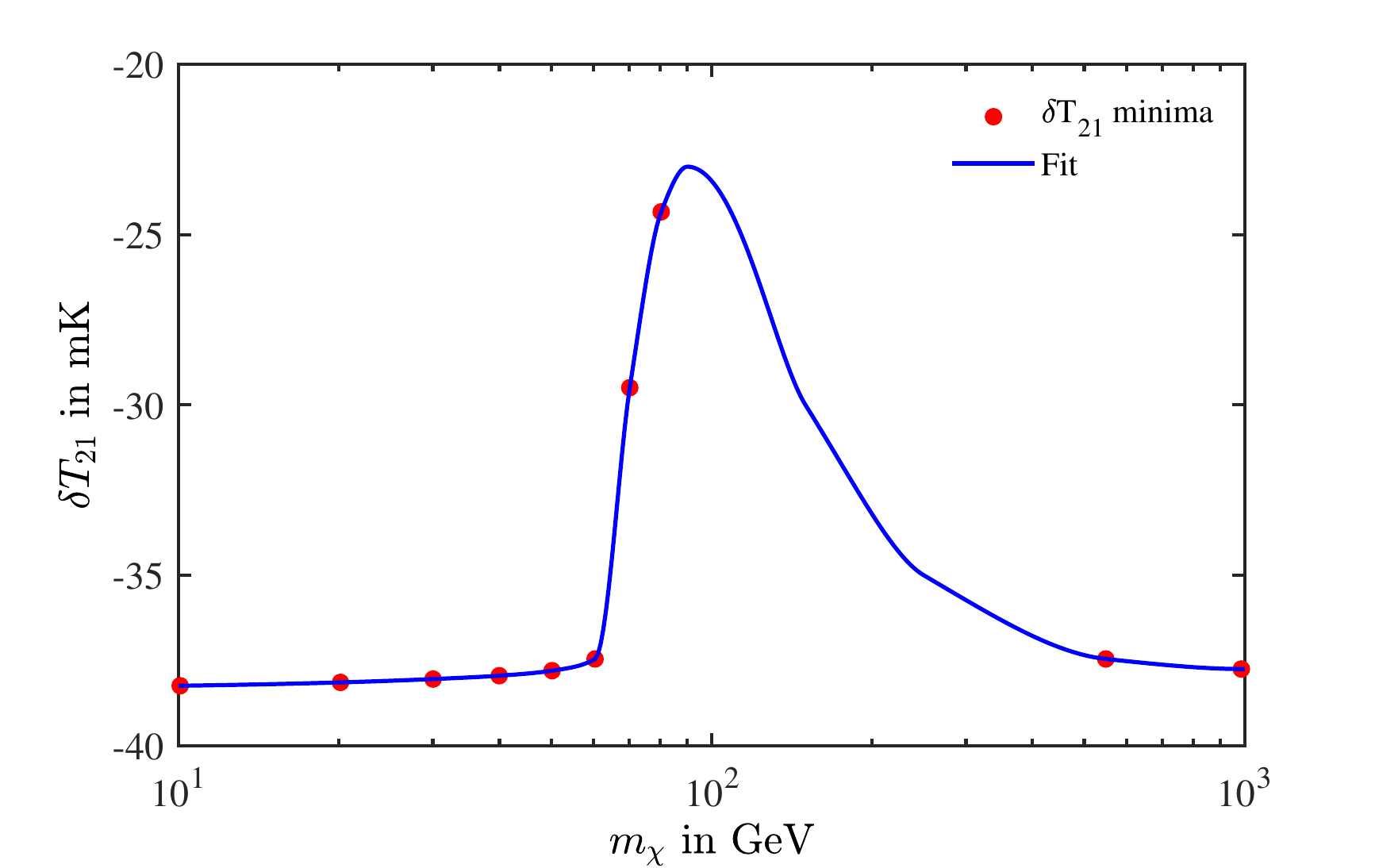}	&
\includegraphics[width=.47\textwidth]{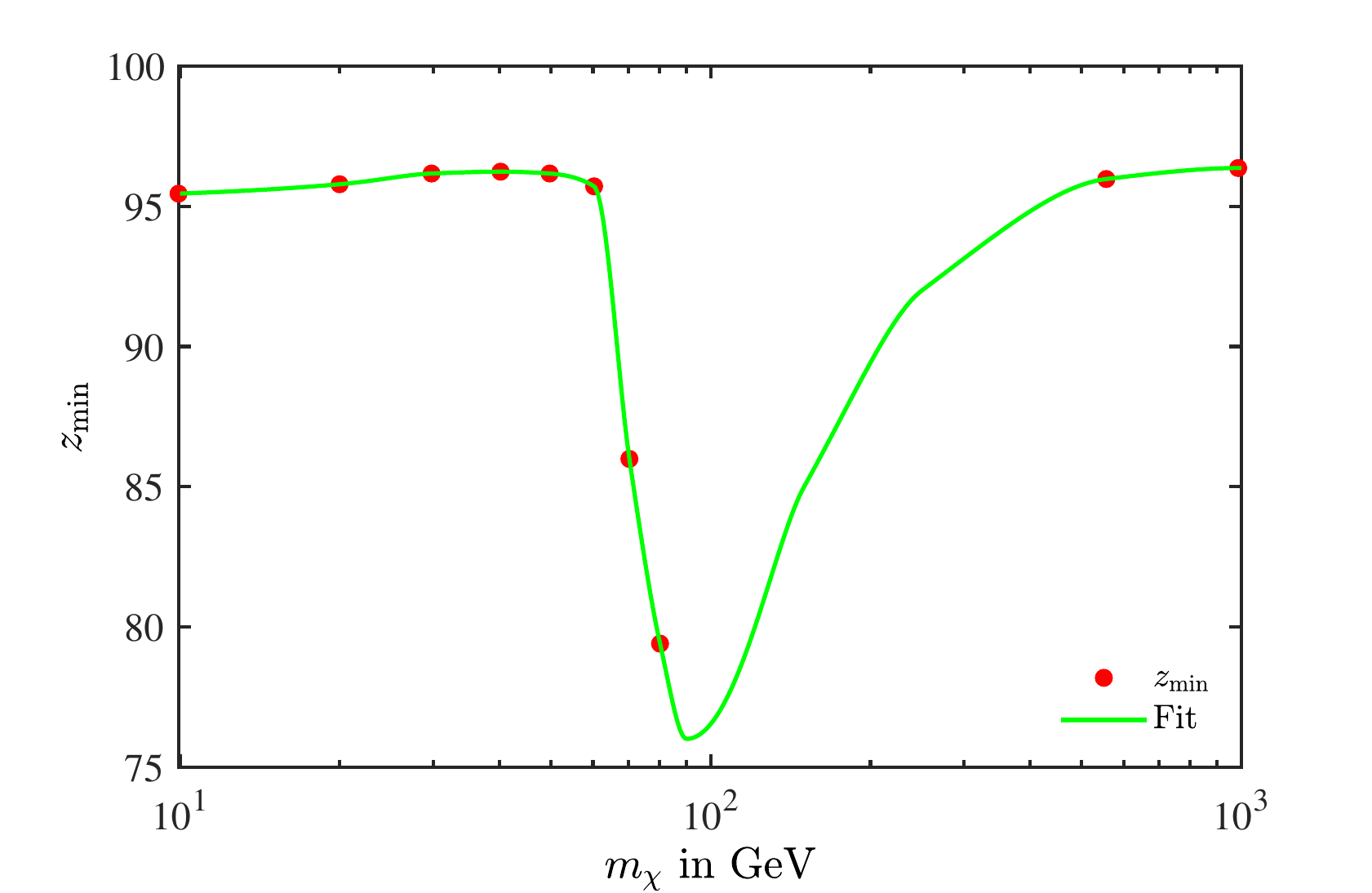} \\
(a) & (b) \\
\end{tabular}}
\caption{(a) The fluctuation in the 21-cm brightness temperature $\delta T_{21}$ at different redshift for combined effects of both 
annihilation and scattering as an additional effect to $\delta T_{21}$ apart from the thermal evolution (red points) along with the fitted curve (blue line). (b) same as (a) but for the variation of 
the redshift values ($z_{\rm min}$), corresponding to each different 
$m_\chi$, at which $\delta T_{21}$ attains minimum  for that $m_\chi$. The solid green line indicates the trends of the variation of $z_{\rm min}$.}
\label{fig:totalfit}
\end{figure}

For better understanding of these findings, we furnish in Fig.~\ref{fig:anifit}(a,b) 
and Fig.~\ref{fig:totalfit}(a,b), the variations of $\delta_T{21}$ minima with $m_\chi$  and the redshift
$z$ at which these minima occur (similar to what is shown 
in Fig.~\ref{fig:colfit}) for the cases when only dark matter self annihilation effects are considered as additional heating effects for the computation of $\delta T_{21}$. Fig.~\ref{fig:totalfit}(a,b) shows similar plots but for the case when both dark matter annihilation and dark matter-baryon scattering are included for the evaluation of $\delta T_{21}$. The solid lines in both Figs.~\ref{fig:anifit} and \ref{fig:totalfit} are not fitted lines but 
indicative lines drawn to assess the possible trends of these variations.
From Fig.~\ref{fig:anifit}(a), it can be observed that while $\delta T_{21}\Big|_{\rm min}$  barely varies for other $m_\chi$ values except for the mass range $\sim 70 - 80$ GeV, when $\delta T_{21}\Big|_{\rm min}$  attains maxima. The corresponding $z_{\rm min}$ values also show similar trend signifying that in this range, $\delta T_{21}$ attains minimum at a redshift value higher than those when dark matter masses ($m_\chi$) are in the range, lower or higher than 
this $\sim 70 - 80$ GeV range for $m_{\chi}$. It can again be noted 
that for this case, $\delta T_{21}$, in general attains minimum for much lower redshifts $z$ than for the case when only dark matter-baryon scattering 
is considered (Figs.~\ref{fig:ani}, \ref{fig:col}). Also for lower $m_\chi$ values, $\delta T_{21}$ is in the ballpark of $-250$ mK, a temperature much lower than $\sim -38$ mK (Fig.~\ref{fig:col}(scattering only)). Similar trend as in Fig.~\ref{fig:totalfit}a is obtained for $\delta T_{21}\Big|_{\rm min}$ for the ``combined" effect of both annihilation and scattering but for $m_\chi$ in the range $\sim 70 - 80$ GeV, the minima of $\delta T_{21}$ are attained at a redshift value lower than those for other chosen dark matter masses.

In our anlysis, we study the effects for the boost factors as mentioned above on brightness temperature $\delta T_{21}$ in the presence of dark matter annihilation and DM-baryon scattering. In several works,\cite{evoli2014unveiling, poulin2015dark, damico18} they have shown that boost factor affects brightness temperature $\delta T_{21}$ due to the additional annihilation effects of the dark matter only in the reionization epoch $z \sim 30$. But in our study we observed that boost factor does not play any significant role at the dark ages as well as in the reionization epoch due to scattering effects along with annihilation effects.

\begin{figure}[h]
\centerline{
\begin{tabular}{@{}ccccc@{}}
\includegraphics[width=.35\textwidth]{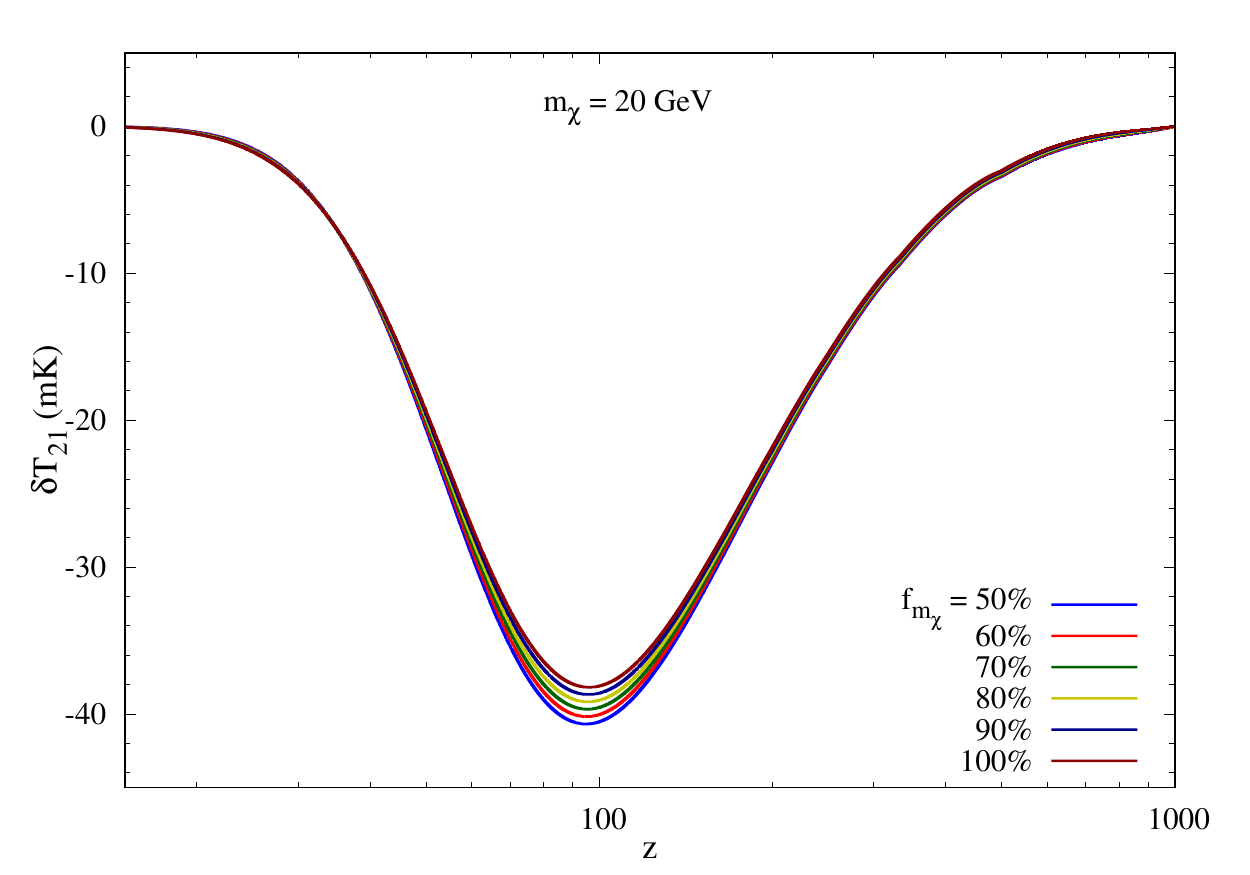}	&
\includegraphics[width=.35\textwidth]{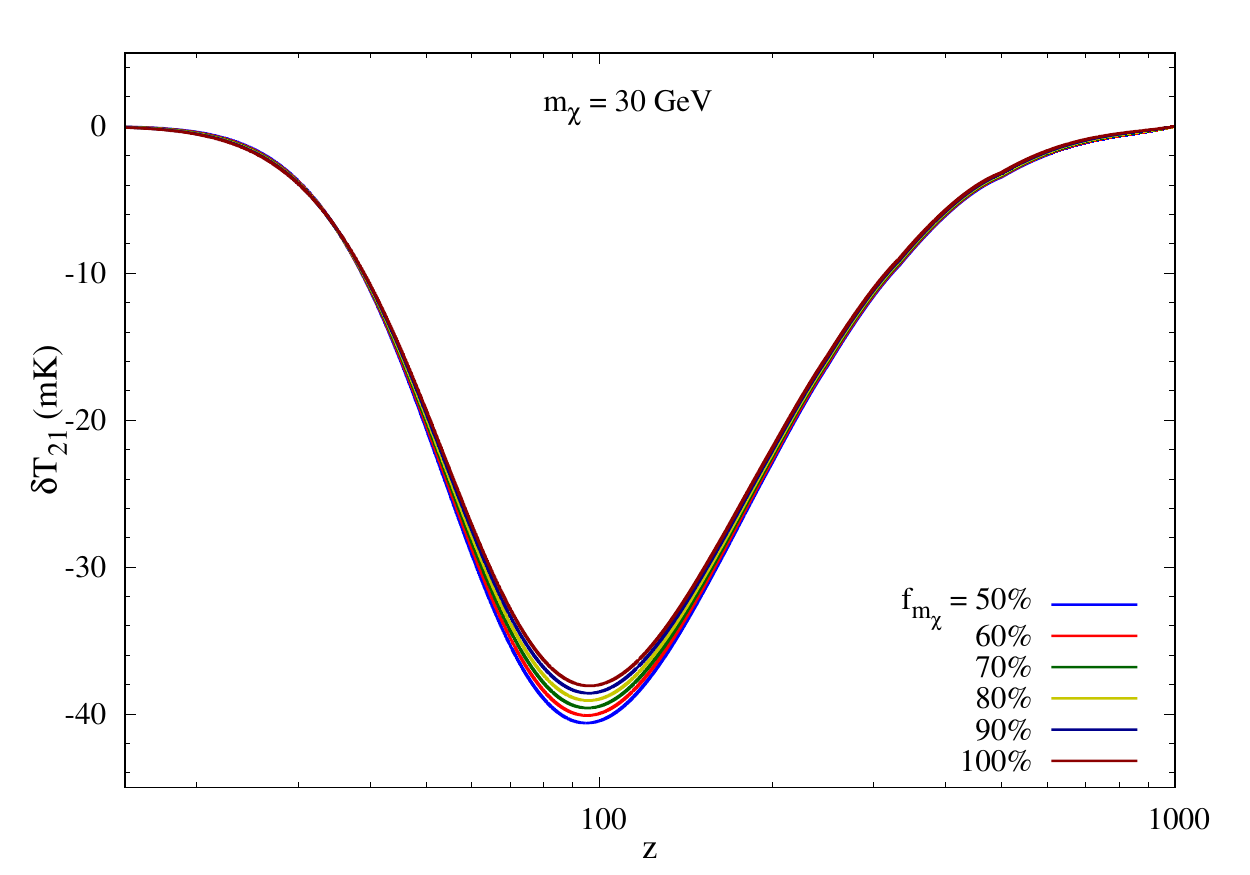} &
\includegraphics[width=.35\textwidth]{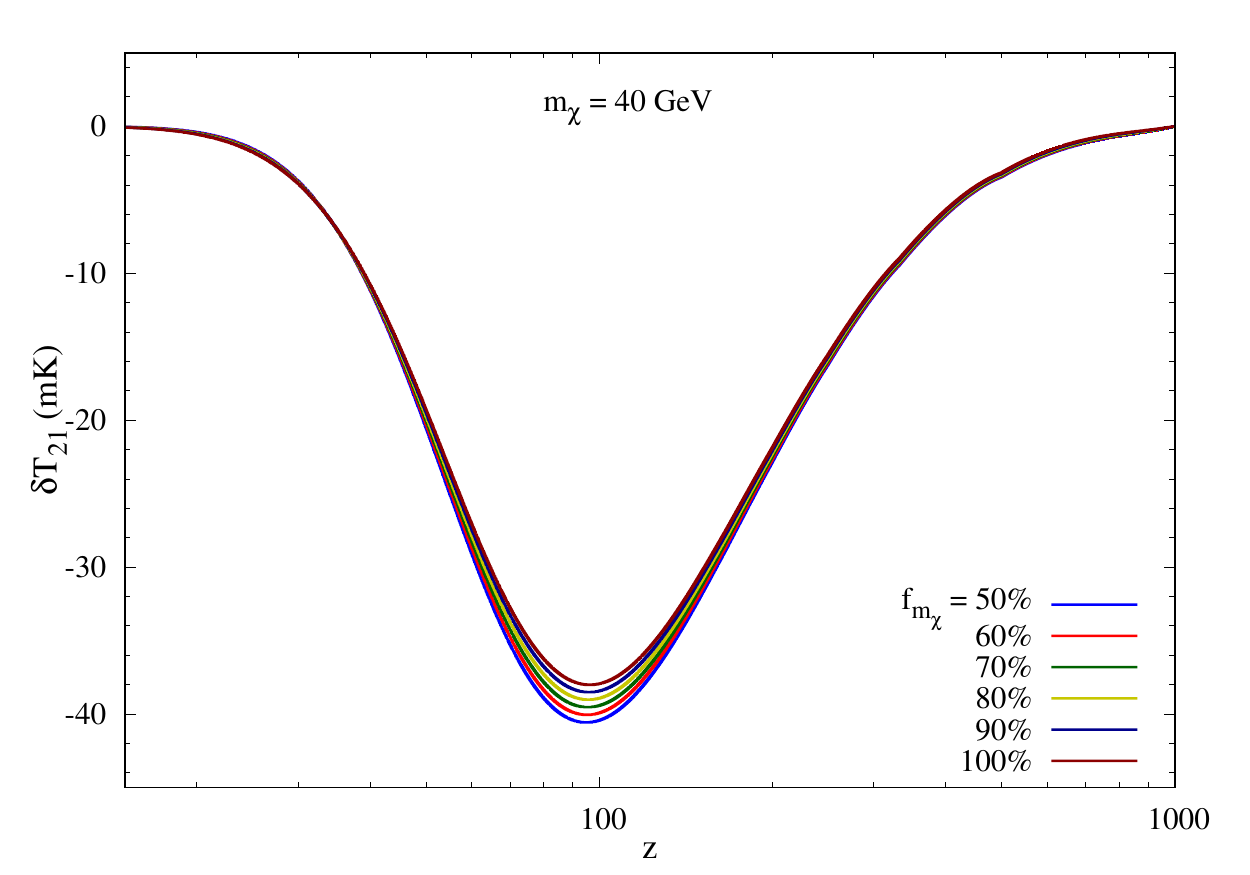}	\\
\includegraphics[width=.35\textwidth]{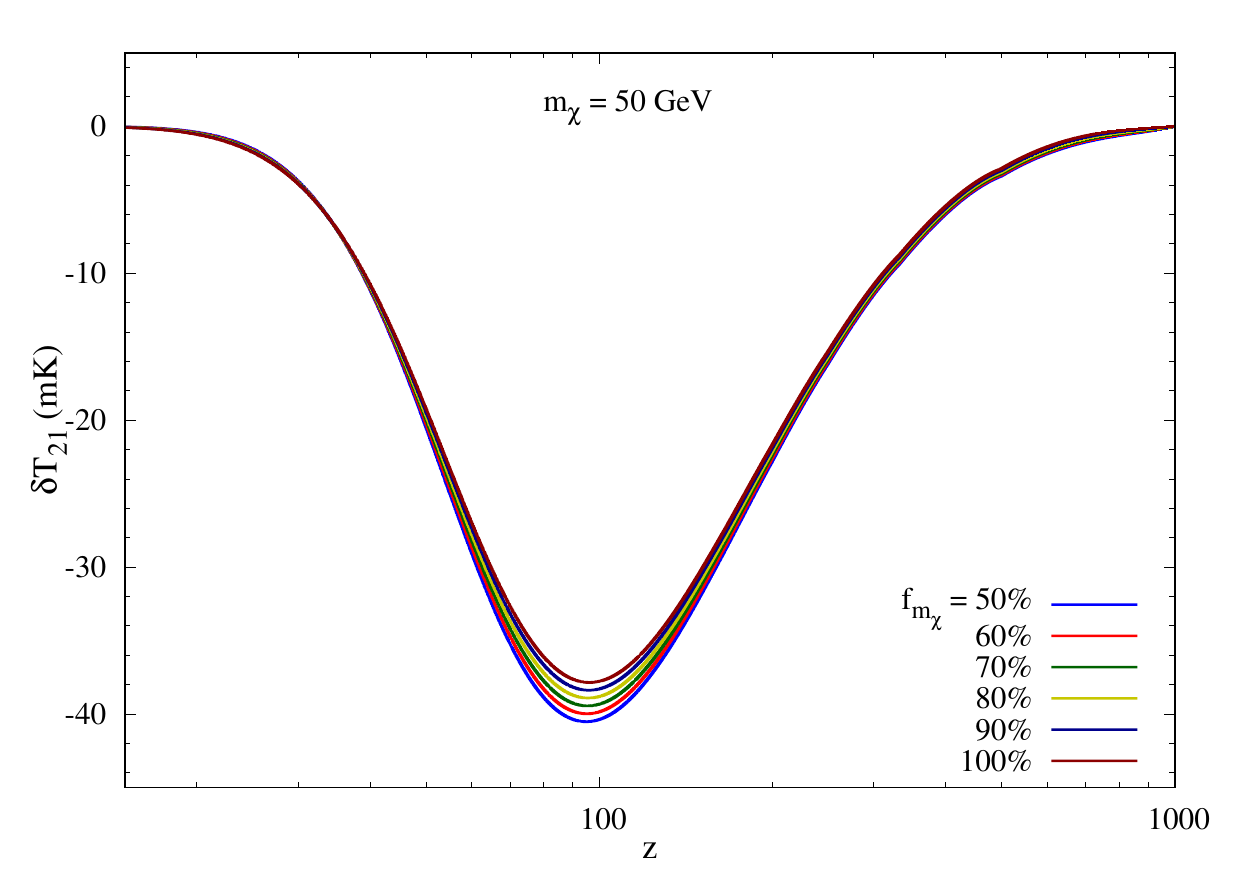}	&
\includegraphics[width=.35\textwidth]{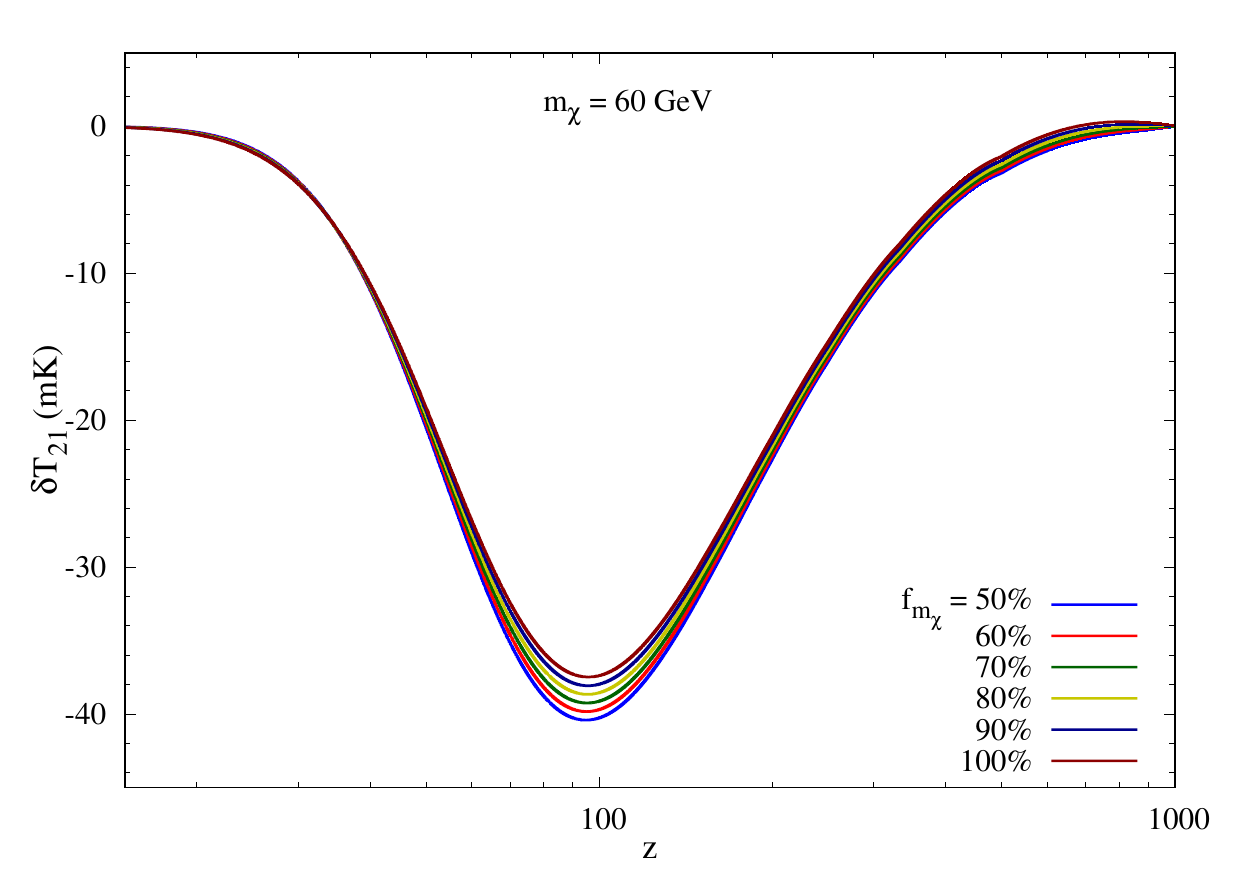}	&
\includegraphics[width=.35\textwidth]{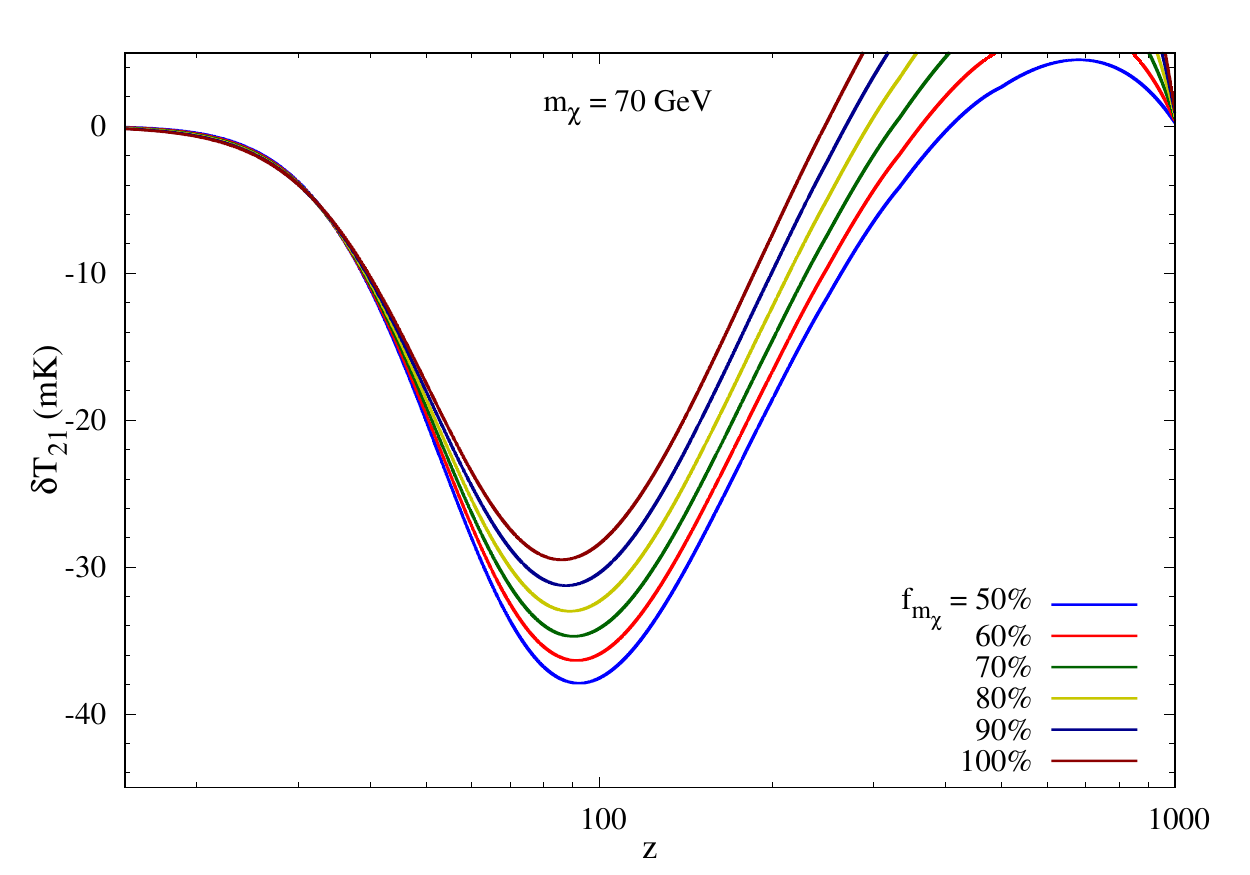}	\\
\includegraphics[width=.35\textwidth]{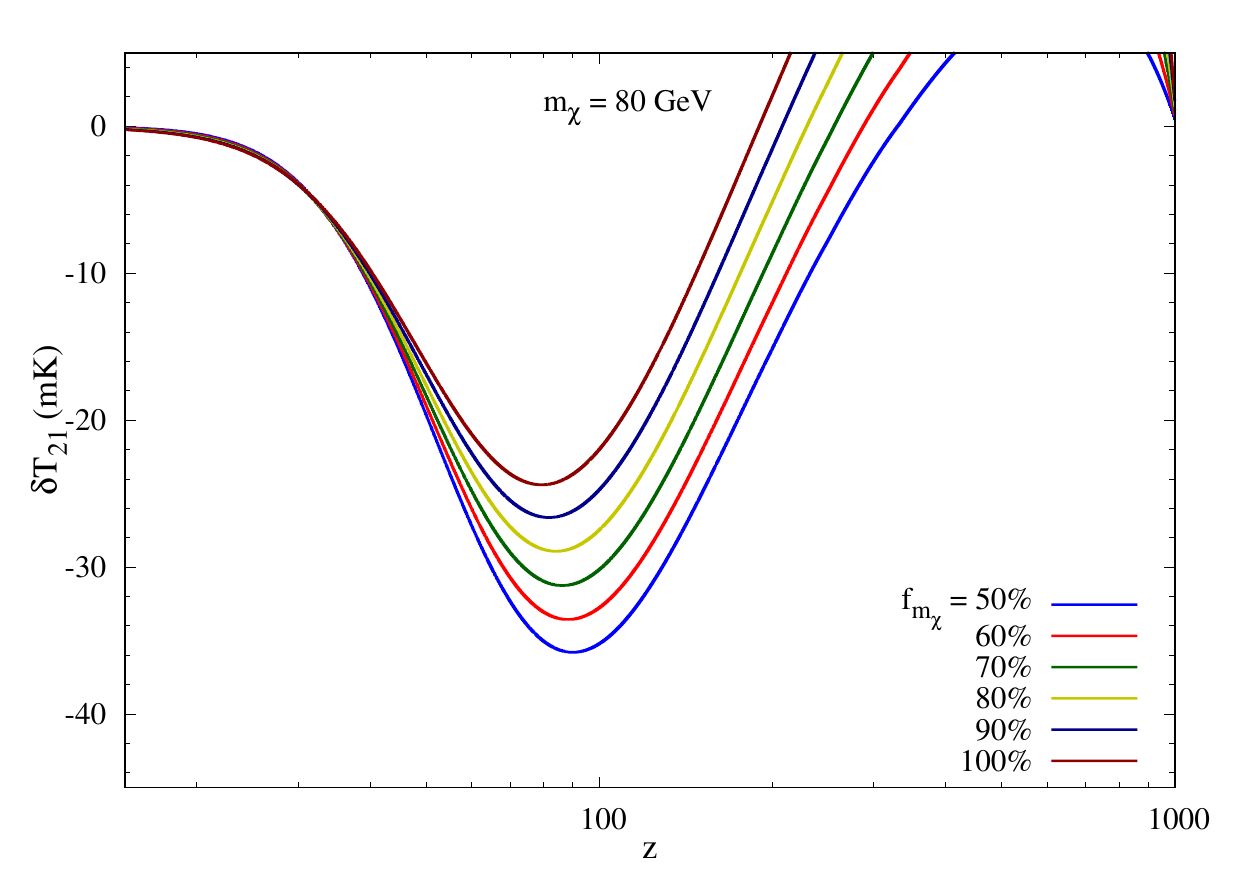}	&
\includegraphics[width=.35\textwidth]{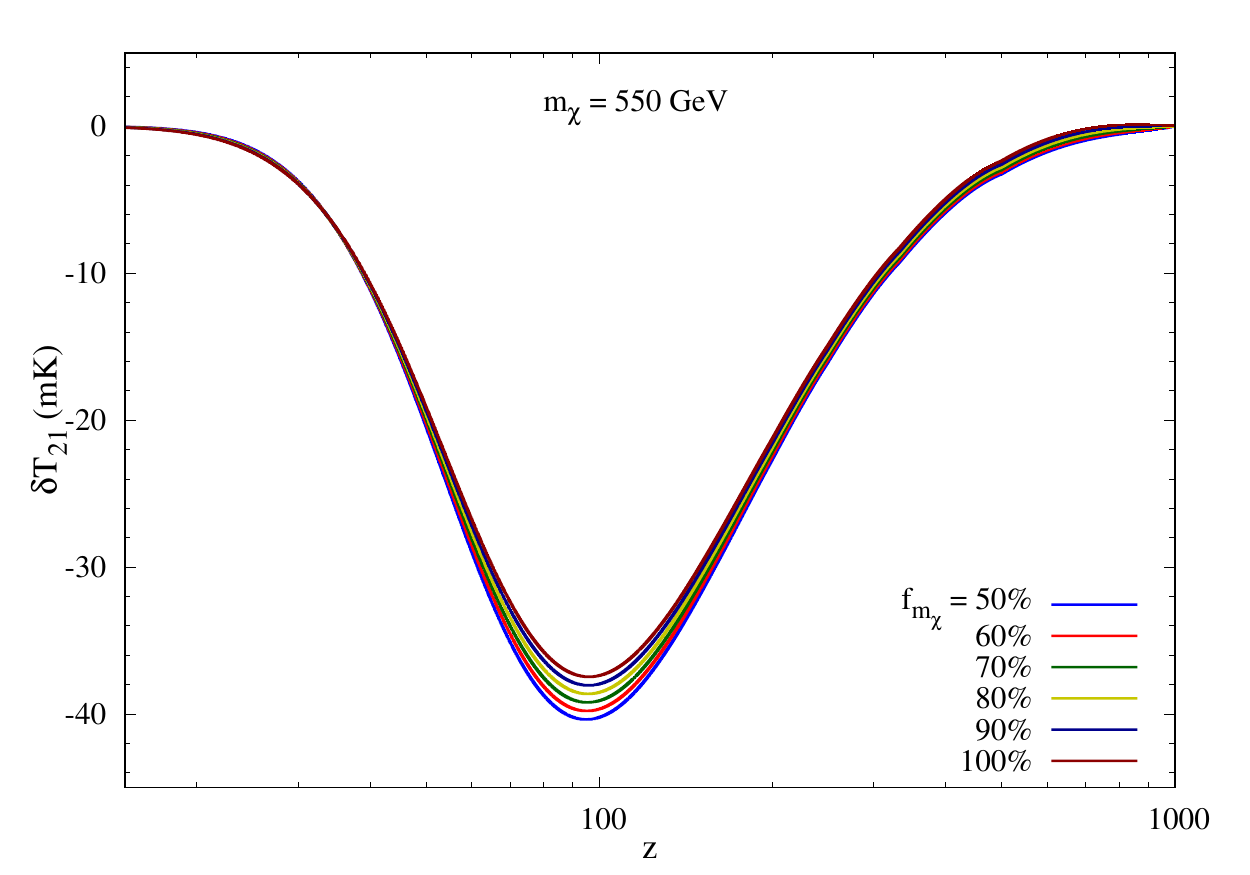} &
\includegraphics[width=.35\textwidth]{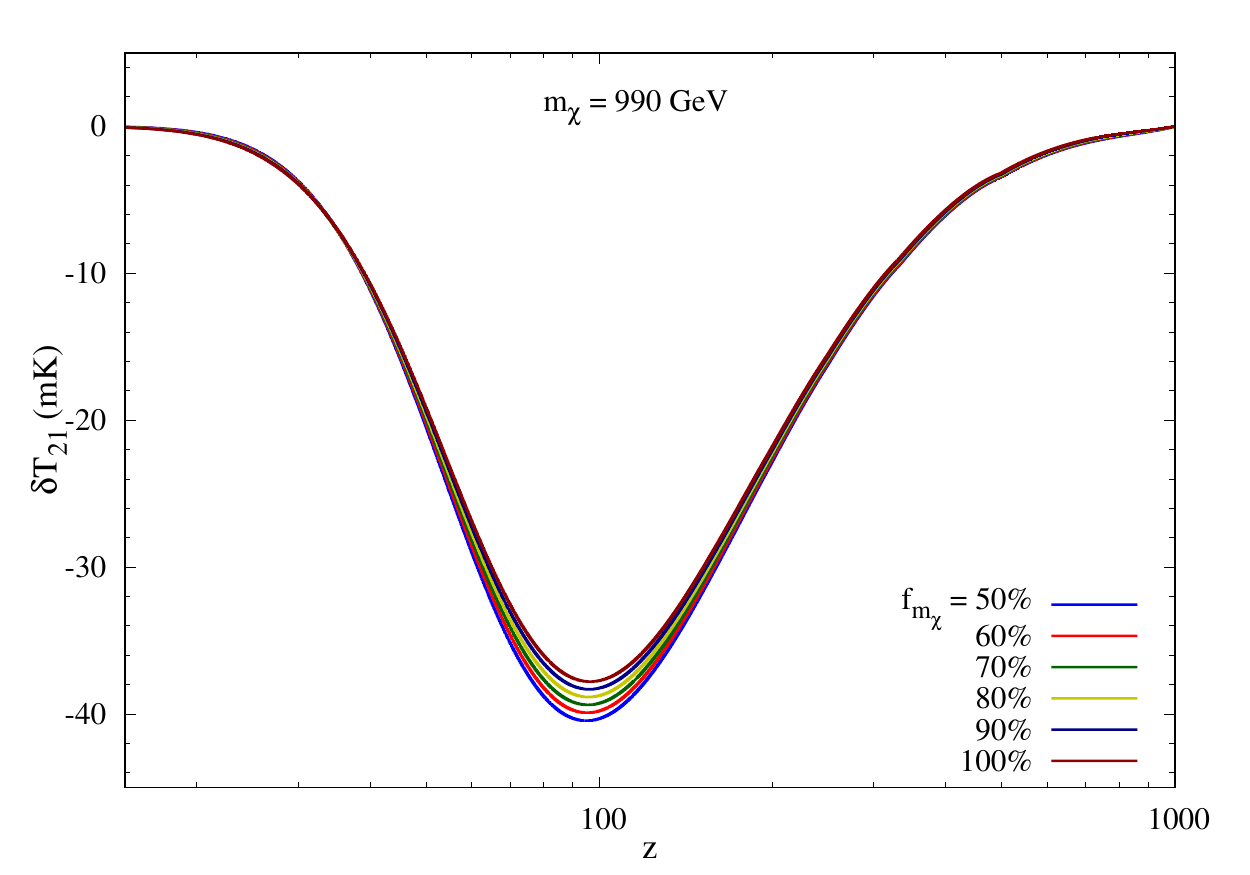} \\
\end{tabular}}
\caption{The fluctuation in the 21-cm absorption lines $\delta T_{21}$ for combined effect in case of different fractions of IDM masses $f_{m_\chi}$.}
\label{fig:multi}
\end{figure}

 We now explore the effect on $\delta T_{21}$, in case the IDM dark matter constitutes
only a fraction of the total dark matter content and only that fraction 
of IDM influences the evolution of $\delta T_{21}$ through their collision 
with baryons and self annihilations. In doing this we consider several 
chosen fractions of IDM dark matter and compute the variations of
$\delta T_{21}$ with $z$ as before. The results for each of the chosen IDM dark
matter masses $m_\chi$ and for a set of chosen fractional values of IDM dark 
matter in the case of each of the chosen $m_\chi$s are shown in Fig.~\ref{fig:multi}. The nine plots in Fig.~\ref{fig:multi} correspond to $m_\chi = 20$, 30, 40, 50, 60, 70, 
80, 550 and 990 GeV and for each of these $m_\chi$s we consider six different
fractions $f_{m_\chi}$ of IDM dark matter namely $f_{m_\chi} = 50$\%, 60\%, 70\%,
80\%, 90\% and 100\%. The features of the plot are similar to those
of Fig.~\ref{fig:total}.

\begin{figure}[h]
\centerline{\includegraphics[scale=0.2]{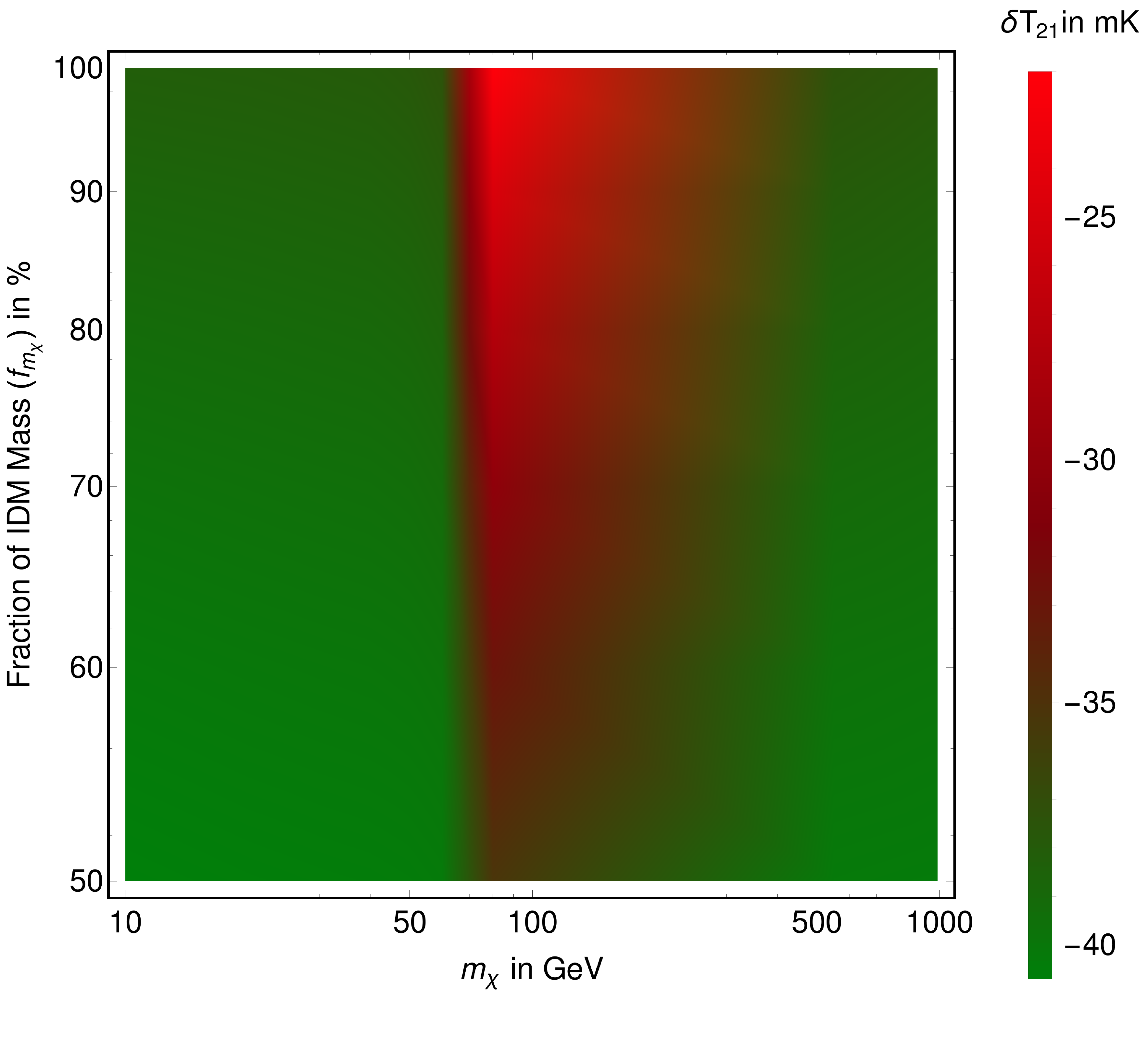}
\includegraphics[scale=0.2]{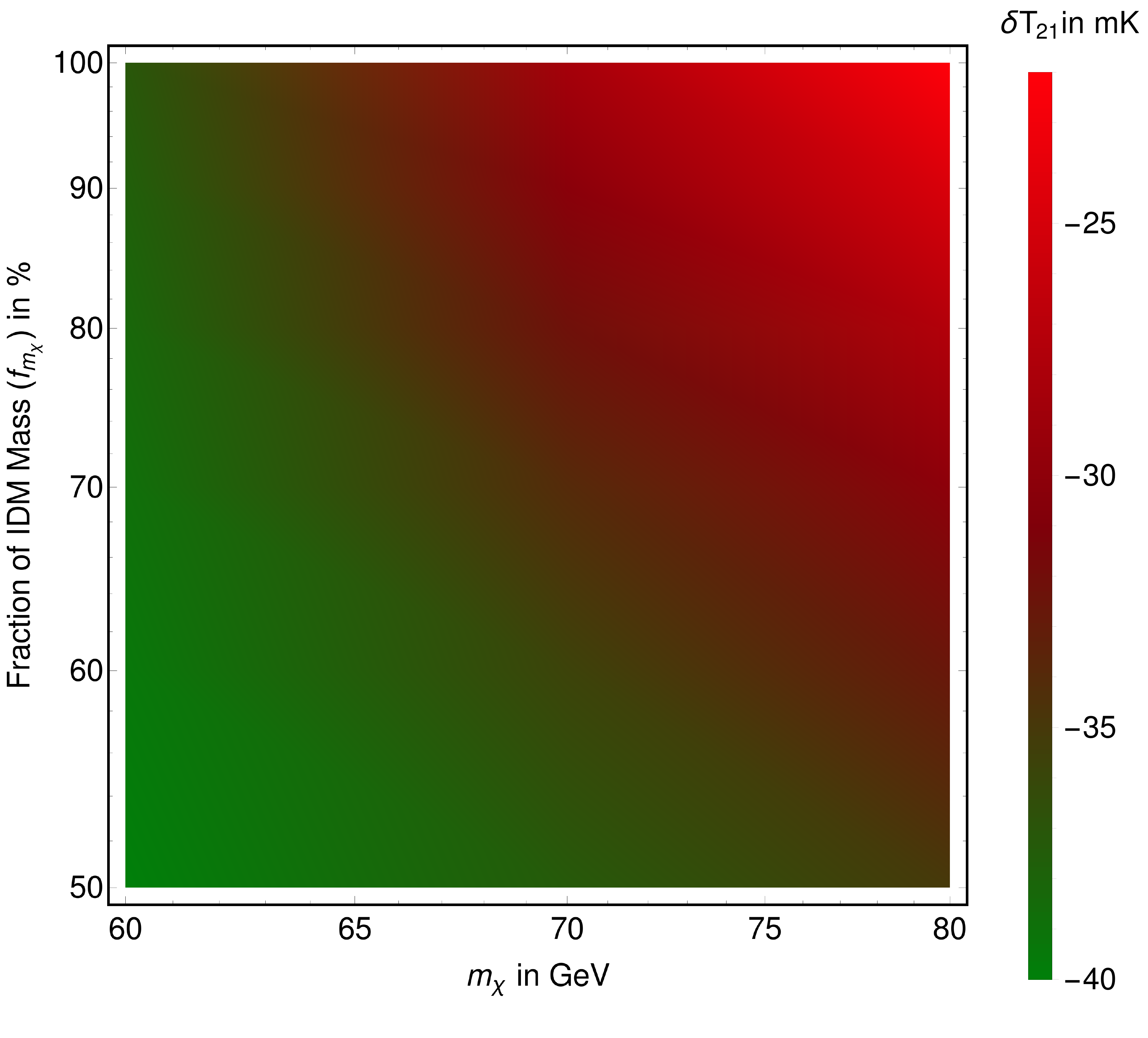}}
\caption{Variations of $\delta T_{21}$ for different fractions of IDM masses $f_{m_\chi}$.}
\label{fig:contour}
\end{figure}

 It can be seen from Fig.~\ref{fig:multi} that $\delta T_{21}$ varies from 
a higher value to lower values as the IDM fraction $f_{m_\chi}$ changes from 
100\% to 50\% for the chosen $m_\chi$s. But except for $m_\chi = 70$, 80
GeV these variations (with the fraction $f_{m_\chi}$) are not very significant.
The $\delta T_{21}\Big|_{\rm min}$ takes lower values as the IDM fraction $f_{m_\chi}$
diminishes. These features are also shown in Fig.~\ref{fig:contour}. Right panel of Fig.~\ref{fig:contour} is similar
to left panel but for a narrower mass range of 60 GeV to 80 GeV for better
understanding.  

\section{Summary and Discussions}\label{sec:4}
With a specific particle dark matter candidate from a proposed particle
physics model, the effect of baryon-dark matter collision and dark 
matter annihilation are addressed for 21cm signal during dark ages leading
to reionization era. The particle dark matter candidate follows from
a SU(2) scalar doublet extended Standard Model of particle physics. The 
extra scalar does not generate any vev (inert) and a $Z_2$ symmetry is imposed 
so that the added scalar is $Z_2$ even. This ensures the stability of 
the inert doublet in this inert doublet model or IDM and it does not have 
direct coupling with SM fermions. The lighter of the neutral scalars 
of the IDM is the candidate for dark matter which is used in this work
to explore the effects on the evolution of 21cm Hydrogen line during the 
dark ages. It is also explored how different fractions of IDM dark matter 
affect the 21cm temperature.

For DM annihilation the energy is deposited to the system either by instantaneous deposition or by delayed deposition. We have seen that for the instantaneous energy deposition, the boost factor become operational, however for the delayed energy deposition there is no such prominent effect of boost factor. Neglecting Wouthuysen-Field effect \cite{hirata06} we have estimated the relation between spin temperature $T_S$ and baryon temperature $T_b$. The evolution in $T_b$ has three major contributions from thermal evolution, DM annihilation and DM-baryon elastic scattering. We have used the formalism prescribed by Ref.~\cite{damico18} to estimate the DM annihilation contribution to $T_b$. Further we have also consider the formalism prescribed by Ref.~\cite{munoz15} to estimate effect of DM-baryon elastic scattering to $T_b$. After estimating $T_b$ we have finally calculated $\delta T_{21}$ for the combined effect of the individual processes mentioned above.

We have used micrOMEGAs code \cite{micromega} in order to estimate the IDM dark matter annihilation cross section and the corresponding relic density for a given IDM dark matter mass. We consider the fact that the relic density of the IDM dark matter should  satisfy the same given by the Planck experiment. In our case relic densities calculated from micrOMEGAs code lie within the 95\% confidence limit of the relic density of dark matter obtained from Planck experiment (see Table \ref{tab:1}).  Also for different fractions of IDM dark matter, we observed that for 100\% IDM dark matter contributions, the dip of $\delta T_{21}$ for combined effect is minimum at $z \sim 95$ ( Fig.~\ref{fig:multi}).  From these, a lower bound of IDM dark matter annihilation cross-section  can be drwan. The lower bound of the annihilation cross section should lie within the range $\langle\sigma v\rangle \sim (6.5 \times 10^{-29})-(4.88\times 10^{-26})\,\, \rm{cm^ 3 / sec}$  for the DM mass range $m_{\chi}\sim 10 - 990 $ GeV. 

The effects due to dark matter-baryon collisions and dark matter annihilations
are addressed separately and the evolution of 21cm signal for 
different dark matter masses are explored for some chosen IDM dark 
matter masses in the range of few tens of GeV and for two chosen 
masses (550 GeV and 990 GeV) in the higher mass region. We then address 
the combined effects of dark matter-baryon collision and dark matter 
annihilation together and compute the time evolution of baryon 
temperature and hence the 21cm temperature $\delta T_{21}$. These 
are achieved by numerically solving a set of coupled differential 
equations with the addition of proper terms in relevant equations
to include these effects. 
The $\delta T_{21}$ is found to be low in lower redshift $z$ region when 
only dark matter annihilation effects are considered. The $\delta_{21}$ 
evolution shows a minimum at low $z$ values (around the epoch of reionization)
for dark matter mass in the region $\sim 70 - 80$ GeV. The effects of 
baryon-dark matter collision appear to be more dominant in low redshift
regions and for all the chosen masses a dip in $\delta T_{21}$ evolution 
is observed around $z \sim 95$. When the combined effect of these two 
features are considered together, the dominance of collision effect 
is seen for $z \sim 100$ beyond which the annihilation effects dominate.  
The effects of different fractions of IDM on $\delta T_{21}$ evolution 
is not very significant for the chosen masses except for IDM dark matter
masses of 70 and 80 GeV. In both the cases, the variations of $\delta T_{21}$
minima with different fractions of IDM dark matter is within $\sim 20 
- 25$\%. It appears that $\delta T_{21}$ is more sensitive in the 
IDM dark matter mass region $\sim 70 - 80$ GeV when the dark matter collision
effects and dark matter annihilation effects are taken into consideration. 

In this work, the Hydrogen 21cm line signature from the dark ages 
leading to the reionization era has been addressed. 
Probing the higher $z$ regions of the dark 
ages would open new vistas in understanding the cosmic evolution and 21cm 
signal. While the EDGES experiment 
explored the redshift region of the reionization era, the experiments
like DAPPER (Dark Ages Polarimeter Pathfinder)\footnote{https://www.colorado.edu/ness/dark-ages-polarimeter-pathfinder-dapper} which is 
to operate in a low lunar orbit, DSL experiment (Discovering the Sky at 
the Longest Wavelengths with Small Satellite Constellation) 
\cite{chen2019discovering} or the Farside Explorer 
\cite{mimoun2012farside} of 
the moon will expect to probe the redshift regions explored in this work.
In addition, the Square Kilometer Array or SKA experiment may 
also look into the dark ages through the radio signal. The data for 21cm
line if available from these experiments in future, will certainly help 
constraining the parameter space
of a particle dark matter model such as the one (IDM model) 
discussed in this work.

\section{Acknowledgements}
Two of the authors (R.B. and S.B.) wish to acknowledge the support received from St. Xavier’s College, Kolkata. One of the authors (R.B.) also thanks the Women Scientist Scheme-A fellowship (SR/WOS-A/PM-49/2018), Department of Science $\&$ Technology (DST), Govt. of India, for providing financial support. One of the author (M.P.) thanks the DST-INSPIRE Fellowship  grant (DST/INSPIRE/FELLOWSHIP/[IF160004]) by DST Govt. of India. We thank  S. Bhattacharyya and A. Halder for some useful discussions and comments.

\end{document}